\documentclass[jgrga]{AGUTeX}

\usepackage{graphicx}
\usepackage{amsmath}
\usepackage{color}

\setkeys{Gin}{draft=false}

\authorrunninghead{HU ET AL.}

\titlerunninghead{MODELING TURBIDITY CURRENTS WITH THE TEM}

\authoraddr{Corresponding author: Zhiguo He, Institute of Physical Oceanography, Ocean College, Zhejiang University, 310058 Hangzhou, China. (hezhiguo@zju.edu.cn)}

\begin{document}


\title{Is it appropriate to model turbidity currents with the three-equation model?}




\authors{Peng Hu,\altaffilmark{1,2}
Thomas P\"ahtz,\altaffilmark{1,2} and
Zhiguo He\altaffilmark{1,2}}

\altaffiltext{1}{Institute of Physical Oceanography, Ocean College, Zhejiang University, 310058 Hangzhou, China.}

\altaffiltext{2}{State Key Laboratory of Satellite Ocean Environment Dynamics, The Second Institute of Oceanography, Hangzhou 310012, China.}



\begin{abstract}

The three-equation model (TEM) was developed in the 1980s to model turbidity currents (TCs) and has been widely used ever since. However, its physical justification was questioned because self-accelerating TCs simulated with the steady TEM seemed to violate the turbulent kinetic energy balance. This violation was considered as a result of very strong sediment erosion that consumes more turbulent kinetic energy than is produced. To confine bed erosion and thus remedy this issue, the four-equation model (FEM) was introduced by assuming a proportionality between the bed shear stress and the turbulent kinetic energy. Here we analytically proof that self-accelerating TCs simulated with the original steady TEM actually never violate the turbulent kinetic energy balance, provided that the bed drag coefficient is not unrealistically low. We find that stronger bed erosion, surprisingly, leads to more production of turbulent kinetic energy due to conversion of potential energy of eroded material into kinetic energy of the current. Furthermore, we analytically show that, for asymptotically supercritical flow conditions, the original steady TEM always produces self-accelerating TCs if the upstream boundary conditions (``ignition'' values) are chosen appropriately, while it never does so for asymptotically subcritical flow conditions. We numerically show that our novel method to obtain the ignition values even works for Richardson numbers very near to unity. Our study also includes a comparison of the TEM and FEM closures for the bed shear stress to simulation data of a coupled Large Eddy and Discrete Element Model of sediment transport in water, which suggests that the TEM closure might be more realistic than the FEM closure.

\end{abstract}

\begin{article}

\section{Introduction}
Turbidity currents (TCs) are sediment-water mixtures rapidly-moving downslope through clear water. They significantly contribute to the evolutions of a large variety of sedimentary structures and morphological features in river reservoirs, lakes, estuaries, and deep oceans \citep{Islametal2008,MeiburgKneller2010,Liuetal2012,Konsoeretal2013} and are thus of high interest for many fields of Earth Science. However, it has turned out quite difficult to carry out controlled in-situ measurements of TCs, explaining why only very few have been reported \citep{Xuetal2004,CossuWells2010,Pylesetal2013,Sumneretal2013}. This highlights the importance of laboratory measurements and numerical modeling of TCs for developing a better understanding of their nature.

There are two groups of numerical models: depth-resolving models \citep{StraussGlinsky2012,Yehetal2013} and layer-averaged models, which include the three-equation model (TEM) and its variants \citep{Fukushimaetal1985,Parkeretal1986,ZengLowe1997,Choi1998,Imranetal1998,BradfordKatopodes1999,KosticParker2006,KosticParker2007,deLunaetal2009,Toniolo2009,HuCao2009,Kosticetal2010,Ekeetal2011,Huetal2012,LaiWu2013,Kostic2014,ElfimovKhakzad2014} and the four-equation-model (FEM) and its variants \citep{Fukushimaetal1985,Parkeretal1986,Salaheldinetal2000,Pratsonetal2001,Dasetal2004,Fildanietal2006,KosticParker2006,YiImran2006,Ekeetal2011,Kostic2011,Traceretal2012}. The FEM differs from the TEM in the way in which the bed shear velocity ($u_\ast$) is computed \citep{Fukushimaetal1985,Parkeretal1986}: while the TEM computes $u_\ast$ from the drag exerted on the bed, roughly approximated by
\begin{eqnarray}
 \mathrm{TEM:}\quad u_\ast^2=C_DU^2, \label{bedshearTEM}
\end{eqnarray}
where $C_D>0$ is the bed drag coefficient and $U$ the layer-averaged velocity of the sediment-water mixture, the FEM computes $u_\ast$ from the assumption that the bed shear stress is proportional to the layer-averaged turbulent kinetic energy ($k$),
\begin{eqnarray}
 \mathrm{FEM:}\quad u_\ast^2=\alpha k, \label{bedshearFEM}
\end{eqnarray}
where $\alpha>0$ is the dimensionless proportionality constant. The inclusion of $k$ in the FEM makes it necessary to also include the turbulent kinetic energy balance. This explains the different names of the models, which suggest that the FEM contains one governing equation more than the TEM. However, in our opinion, these names are slightly misleading since the same turbulent kinetic energy balance can also be used in TEM to compute $k$ \citep{Fukushimaetal1985,Parkeretal1986}. The actual difference is that $k$ influences the evolution of TCs in the FEM, while it does not do so in the TEM.

Why were two kinds of layer-averaged models, the TEM and FEM, developed for TCs? The answer is that the steady TEM simulations by \citet{Fukushimaetal1985} (F85) and \citet{Parkeretal1986} (P86) failed to reproduce physically realistic self-accelerating TCs, which occur when the bed slope ($S$) is sufficiently large to ensure that $U$ and the sediment transport rate ($\psi$) increase downstream without limit ever after a sufficiently large, finite distance downstream ($\Leftrightarrow(U,\psi)\xrightarrow{x\rightarrow\infty}(\infty,\infty)$, where $x$ is the streamwise coordinate). In fact, the authors found that the dimensionless net production rate of the turbulent kinetic energy ($\Delta E=(h/U^2)dk/dx$, where $h$ is the depth of the TC), composed of production through conversion from potential energy and dissipation through erosion and suspension of bed sediment, becomes negative when $x\rightarrow\infty$ for self-accelerating TCs. It follows that the TCs should die, which is inconsistent with its self-accelerating property \citep{Fukushimaetal1985,Parkeretal1986}. The authors linked this ostensible failure of the steady TEM to a possible overestimation of the bed sediment erosion rate, caused by a possible overestimation of $u_\ast$ in Eq.~(\ref{bedshearTEM}), and remedied this issue by assuming Eq.~(\ref{bedshearFEM}), which limits $u_*$ through a simplified first-order relation with $k$.

In this paper, we report simulations of self-accelerating TCs using the steady TEM and parameter values and empirical relations exactly as reported by F85 and P86. We find that the simulations by P86 do not produce self-accelerating, but instead decelerating TCs when the upstream boundary conditions (``ignition values'') specified by P86 are used. The authors obtained these ignition values by setting $h(0)=2$m, and following the procedure by \citet{Parker1982}. However, if instead $h(0)=1$m is used, the simulations by P86 do result in physically realistic self-accelerating TCs, in contrast to the claim made in this study that such TCs do not occur for the specified parameter range. Moreover, we also find that simulations by F85 do result in physically realistic self-accelerating TCs even for the same ignition values, in contrast to the claim made in this study. In fact, we find that the downstream profile of $dk/dx$ is actually exactly opposite to the one described in F85, and thus $\overrightarrow{\Delta E}>0$ (where the arrow denotes hereafter the limit $x\rightarrow\infty$, $\overrightarrow{\,\cdot\,}=\lim_{x\rightarrow\infty}\cdot$), \textit{consistent} with its self-accelerating property, indicating a possible error in the computations by F85. We support this claim with an analytical proof showing that $\overrightarrow{\Delta E}>0$ for self-accelerating TCs simulated with the steady TEM if $C_D$ or alternatively $S$ are not unrealistically low. This proof contains the derivation of the asymptotic behaviors ($x\rightarrow\infty$) of quantities characterizing a self-accelerating TC, such as the Richardson number ($Ri$), which we show to be asymptotically constant. From an analytical stability analysis of the steady TEM, we then find that $\overrightarrow{Ri}<1$ (supercritical flow, i.e., the densimetric Froude number $Fr>1$ since $Ri=1/Fr^2$) is a necessary and sufficient condition for the existence of self-accelerating TCs. On basis of this result, we provide a novel method to find ignition values which always result in self-accelerating TCs for simulations using the steady TEM. Another interesting finding of our study is that, surprisingly, a larger bed sediment erosion rate, $E_s$, results in larger values of $\Delta E$, even though the erosion of bed sediment dissipates turbulent kinetic energy. This is because eroded bed sediment increases the sediment mass and thus potential energy of the TC, which is then converted into turbulent kinetic energy downslope. Our study also includes a comparison of the TEM and FEM closures for the bed shear stress (Eqs.~(\ref{bedshearTEM}) and (\ref{bedshearFEM})) to simulation data of a coupled Large Eddy and Discrete Element Model of sediment transport in water \citep{FurbishSchmeeckle2013,Schmeeckle2014}, which suggests that the TEM closure might be more realistic than the FEM closure.

In the following, we first briefly review the conservation equations governing the steady TEM and FEM as reported by F85 and P86 in Section \ref{Conserveq}. Then we proof that $\overrightarrow{\Delta E}>0$ if $C_D$ and $S$ are not unrealistically low for self-accelerating TCs simulated with the steady TEM in Section \ref{Proof}. This section also contains the proof that $\overrightarrow{Ri}<1$ is a necessary and sufficient condition for the existence of self-accelerating TCs. Afterwards in Section \ref{simulations}, we present our simulations using the steady TEM and parameter values exactly as reported in F85 and P86 and show that these simulations are consistent with our analytical proof. There we also present a method to find ignition values which always result in self-accelerating TCs for simulations using the steady TEM and show that stronger erosion leads to larger positive values of $\Delta E$. The latter is then explained in Section \ref{Discussion}, which also includes the comparison of the TEM and FEM closures for the bed shear stress to the aforementioned numerical data, and conclude in Section \ref{Conclusion}.

\section{Conservation equations} \label{Conserveq}
Although unsteady models might be more realistic, we here consider steady TCs ($\partial/\partial t=0$) in order to be consistent with the original studies by F85 and P86. For this case, the mass conservations of the sediment-water mixture (Eq. (\ref{massmixture})) and the sediment carried by the current (Eq. (\ref{momentummixture})), the momentum conservation of the sediment-water mixture (Eq. (\ref{masssediment})), and the turbulent kinetic energy conservation of the sediment-water mixture (Eq. (\ref{energymixture})) are written as \citep{Fukushimaetal1985,Parkeretal1986}
\begin{eqnarray}
 \frac{dh}{dx}&=&\frac{-RiS+e_w(2-0.5Ri)+\frac{u_\ast^2}{U^2}+\frac{1}{2}RiR_\psi}{1-Ri}, \label{massmixture} \\
 \frac{h}{U}\frac{dU}{dx}&=&\frac{RiS-e_w(1+0.5Ri)-\frac{u_\ast^2}{U^2}-\frac{1}{2}RiR_\psi}{1-Ri}, \label{masssediment} \\
 \frac{h}{\psi}\frac{d\psi}{dx}&=&\frac{v_s}{U}r_o\left(\frac{\psi_e}{\psi}-1\right)=R_\psi, \label{momentummixture} \\
 \frac{h}{U^2}\frac{d k}{dx}&=&\left(\frac{1}{2}e_w(1-Ri)+\frac{u_\ast^2}{U^2}-\frac{ke_w}{U^2}\right. \nonumber \\
 &-&\left.\frac{\beta k^{3/2}}{U^3}-Ri\frac{v_s}{U}-\frac{1}{2}RiR_\psi\right)\Theta(k)=\Delta E, \label{energymixture}
\end{eqnarray}
where $\Theta$ denotes the Heaviside function, $Ri=\tilde g\psi/U^3$ the Richardson number with $\tilde g=g(\rho_s-\rho_w)/\rho_w>0$ the submerged value of the gravity constant ($g$) and $\rho_s$ ($\rho_w$) the density of sediment (water), $v_s>0$ is the sediment settling velocity,
\begin{eqnarray}
 \psi_e=E_shU/r_o \label{psie}
\end{eqnarray}
is the equilibrium sediment transport rate at which sediment exchange between the current and the bed vanishes ($R_\psi=0$), while $e_w\geq0$ (water entrainment rate), $r_o>1$ (ratio between bed and average sediment concentration), $E_s\geq0$ (bed sediment erosion rate), and $\beta=he_o/k^{3/2}>0$ ($e_o>0$ is the average rate of viscous dissipation of turbulent kinetic energy) are bounded coefficients. We note that Eq.~(\ref{energymixture}) has been slightly modified from the version reported by F85 and P86, namely it has been multiplied by $\Theta(k)$. While this modification has no relevance for practical applications because $\Theta(k)=1$ if $k>0$, we incorporated it here for the mathematical proof in Section~\ref{Proof} since it ensures that $\Delta E$ does not become a complex number due to $\Theta(k)=0$ if $k\leq0$. Indeed, without $\Theta(k)$, $k$ could become negative and thus $\Delta E$ complex due to the term $k^{3/2}$. Eqs.~(\ref{massmixture}-\ref{momentummixture}) constitute the steady TEM together with the closure Eq.~(\ref{bedshearTEM}). In contrast, the steady FEM is constituted by Eqs.~(\ref{massmixture}-\ref{energymixture}) since the closure Eq.~(\ref{bedshearFEM}) incorporates $k$, which must be computed by Eq.~(\ref{energymixture}). However, even though Eq.~(\ref{energymixture}) does not influence the values of $h$, $U$, and $\psi$ in the steady TEM, it is still used to compute $\Delta E$ if required \citep{Fukushimaetal1985,Parkeretal1986}. Moreover, it is important to point out the fact that the computed layer-averaged volumetric sediment concentration, defined by
\begin{eqnarray}
 C=\frac{\psi}{hU}, \label{conc}
\end{eqnarray}
must be smaller than unity, which is automatically ensured by the steady TEM for most practical applications (mainly due to Eq.~(\ref{momentummixture}), which makes $\psi$ decrease strongly when $C$ becomes large). In fact, as we show in Section~\ref{Proof}, Eq.~(\ref{conc}) is always fulfilled for self-accelerating TCs in the limit $x\rightarrow\infty$.

F85 and P86 used the following empirical relationships to compute the coefficients $e_w$, $r_o$, $E_s$ in the steady TEM and FEM and $\beta$ in the FEM,
\begin{eqnarray}
 e_w&=&0.00153/(0.0204+Ri), \label{ew} \\
 r_o&=&1+31.5(u_\ast/v_s)^{-1.46}, \\
 E_s&=&\left\{ 
  \begin{array}{l l}
    0.3 & Z\geq13.2 \\
    3\times10^{-12}Z^{10}(1-5/Z) &  5<Z<13.2 \\
    0 & Z\leq5
  \end{array} \right\}, \label{Es} \\
 \beta&=&\frac{0.5e_w(1-Ri-2C_D/\alpha)+C_D}{(C_D/\alpha)^{1.5}}\quad\mathrm{(only\;FEM)}, \label{beta}
\end{eqnarray}
where $Z=\sqrt{Re_p}(u_\ast/v_s)$ with $Re_p=\sqrt{\tilde gD_s^3}/\nu$ the particle Reynolds number, $D_s$ the mean sediment particle diameter, and $\nu$ the kinematic viscosity of clear water. From simulations with the steady TEM using Eqs.~(\ref{ew}-\ref{Es}) and $S=0.08$, F85 obtained $\overrightarrow{\Delta E}<0$ for their simulated self-accelerating TCs and thus concluded that the TEM produces physically unrealistic results. In the following section, we analytically show that the steady TEM can only result in $\overrightarrow{\Delta E}<0$ if $C_D$ and $S$ are both smaller than certain threshold values defined later. In particular, if Eq.~(\ref{ew}) is used to compute $e_w$, $C_D$ must be smaller than $0.00097$ (which is much smaller than the authors' $C_D=0.004$) and $S$ be smaller than $0.0073$ (which is much smaller than the authors' $S=0.08$).

\section{Analytical proof} \label{Proof}
In this section, we proof that $\overrightarrow{\Delta E}>0$ for self-accelerating TCs simulated with the steady TEM if $C_D\geq C_{D{\mathrm{min}}}$ or $S\geq S_{\mathrm{min}}$, where $C_{D\mathrm{min}}$ and $S_\mathrm{min}$ are certain values of $C_D$ and $S$, respectively, which we define later. We also proof that $\overrightarrow{Ri}<1$ is a necessary and sufficient condition for the existence of self-accelerating TCs. For the proof, we make use of the definition of self-accelerating TCs, which includes the property $\overrightarrow{U}=\overrightarrow{\psi}=\infty$. Our proof does not require particular empirical expressions or values for the empirical parameters $C_D$, $e_w$, $r_o$, $E_s$, and $\beta$, such as Eqs.~(\ref{ew}-\ref{beta}). It only requires that these parameters are bounded as well as $\lim_{U\rightarrow\infty}E_s>0$ (bed material must be eroded if $U$ is sufficiently large), $e_w(Ri<\infty)>0$, $e_w\xrightarrow{Ri\rightarrow\infty}0$ (water is entrained if and only if flow turbulence is present), and $de_w/dRi\leq0$ (water entrainment increases with turbulence). All properties needed for the proof are summarized in Table~\ref{premisses}.
\begin{table}
\caption{Summary of properties needed for the analytical proof}
\centering
  To show: $\overrightarrow{U}=\overrightarrow{\psi}=\infty\Rightarrow\overrightarrow{\Delta E}>0$ if $C_D\geq C_{D{\mathrm{min}}}$ or $S\geq S_{\mathrm{min}}$ \\
\begin{tabular}{|c|c|}
\hline
  Property & Explanation   \\
\hline
 Eqs.~(\ref{bedshearTEM}) and (\ref{massmixture}-\ref{energymixture}) & definition of the TEM \\
 $\overrightarrow{h}>0$, $\overrightarrow{U}=\overrightarrow{\psi}=\infty$ & definition of self-accelerating current \\
 $0<\tilde g=\overrightarrow{\tilde g}<\infty$ & constant, positive parameter \\
 $0<S=\overrightarrow{S}<\infty$ & constant, positive parameter \\
 $0<C_D=\overrightarrow{C_D}<\infty$ & constant, positive parameter \\
 $0<v_s=\overrightarrow{v_s}<\infty$ & constant, positive parameter \\
 $1\leq\overrightarrow{r_o}<\infty$ & $r_o>1$ is a bounded parameter \\
 $0\leq\overrightarrow{\beta}<\infty$ & $\beta>0$ is a bounded parameter \\
 $0\leq\overrightarrow{e_w}<\infty$ & $e_w\geq0$ is a bounded parameter \\
 $0\leq\overrightarrow{E_s}<\infty$ & $E_s\geq0$ is a bounded parameter \\
 $\lim_{U\rightarrow\infty}E_s>0$ & erosion occurs for infinite current speed \\
 $e_w(Ri<\infty)>0$ & water is entrained by turbulence \\
 $\lim_{Ri\rightarrow\infty}e_w=0$ & no entrainment without turbulence \\
 $de_w/dRi\leq0$ & entrainment increases with turbulence \\
\hline
\end{tabular}
\label{premisses}
\end{table}

Moreover, in our proof we often formally operate with quantities in the limit $x\rightarrow\infty$, which requires that this limit exists for these quantities (note that a limit also exists, if it is infinite). However, functions with spatial periodicity might not fulfill this requirement (e.g., $\overrightarrow{\sin x}$ does not exist). Since Eqs.~(\ref{massmixture}-\ref{energymixture}) do not explicitly contain periodic functions, it is safe to presume that such limits always exist for our case. Finally, we will often use the following rewritten versions of Eq.~(\ref{momentummixture}),
\begin{eqnarray}
 R_\psi&=&E_sv_sh/\psi-r_ov_s/U, \label{Rpsihilf1} \\
 \overrightarrow{R_\psi}&=&\overrightarrow{E_s}v_s\overrightarrow{h/\psi}, \label{Rpsihilf3} \\
 d\psi/dx&=&E_sv_s-r_ov_sC, \label{Rpsihilf2} \\
 \overrightarrow{d\psi/dx}&=&\overrightarrow{E_s}v_s-\overrightarrow{r_o}v_s\overrightarrow{C}, \label{Rpsihilf4}
\end{eqnarray}
where we used Eqs.~(\ref{psie}) and (\ref{conc}), and $\overrightarrow{U}=\infty$, and that $\overrightarrow{E_s}>0$ (from $\lim_{U\rightarrow\infty}E_s>0$), $\overrightarrow{E_s}<\infty$, $0<v_s<\infty$, and $0<\overrightarrow{r_o}<\infty$. Eqs.~(\ref{Rpsihilf3}) and (\ref{Rpsihilf4}) are the limits $x\rightarrow\infty$ of Eqs.~(\ref{Rpsihilf1}) and (\ref{Rpsihilf2}), respectively.

Our proof is separated into three parts. First, we calculate $\overrightarrow{R_\psi}$, $\overrightarrow{Ri}$, and the asymptotic behaviors of $U$, $\psi$, and $h$ in Section~\ref{bounded}. Using the results of Section~\ref{bounded}, we then proof in Section~\ref{existence} that $\overrightarrow{Ri}<1$ is a necessary and sufficient condition for the existence of self-accelerating TCs, and in Section~\ref{necessary} that $C_D\geq C_{D{\mathrm{min}}}$ or $S\geq S_{\mathrm{min}}$ are sufficient conditions for $\overrightarrow{\Delta E}>0$. Afterwards we discuss why these condition are virtually always fulfilled for physically relevant cases in Section~\ref{relevantcases}.

\subsection{Asymptotic solutions} \label{bounded}
This part of the proof follows several logically ordered steps to show $0<\overrightarrow{Ri}<\infty$ and $\overrightarrow{R_\psi}<\infty$. (Note that $\overrightarrow{Ri}>0$ is not trivial, even though $Ri>0$ is fulfilled, since in the limit this quantity could approach the boundaries of its restricted domain.) Afterwards we use these results to calculate $\overrightarrow{R_\psi}$, $\overrightarrow{Ri}$ and the asymptotic behaviors of $U$, $\psi$, and $h$. Note that all results of this section are formally also valid for the FEM if one replaces $C_D$ by $\overrightarrow{\alpha k/U^2}$ (from Eqs.~(\ref{bedshearTEM}) and (\ref{bedshearFEM})) and assumes $0<\overrightarrow{\alpha k/U^2}<\infty$.

\subsubsection{Showing $\overrightarrow{Ri}>0$}
$\overrightarrow{Ri}>0$ can be shown by presuming $\overrightarrow{Ri}=0$ and arriving at a contradiction. Inserting this presumption in the limit $x\rightarrow\infty$ of Eq.~(\ref{masssediment}) using Eq.~(\ref{bedshearTEM}) yields
\begin{eqnarray}
 \overrightarrow{\frac{h}{U}\frac{dU}{dx}}=-\overrightarrow{e_w}-C_D-\frac{1}{2}\overrightarrow{RiR_\psi}<0, \label{help3}
\end{eqnarray}
where we used $\overrightarrow{e_w}\geq0$, $C_D>0$, and $\overrightarrow{RiR_\psi}\geq0$ (from $\overrightarrow{U}=\overrightarrow{\psi}=\infty$). Eq.~(\ref{help3}) is a contradiction to $\overrightarrow{U}=\infty$, which requires $\overrightarrow{dU/dx}\geq0$ and thus $\overrightarrow{\frac{h}{U}\frac{dU}{dx}}\geq0$. Hence, $\overrightarrow{Ri}>0$.

\subsubsection{Showing $\overrightarrow{C}<\infty$}
While it is physically clear that $C$ cannot be larger than unity, we do not need to assume this beforehand for the analytical proof. Instead, this property is strictly obtained from the model equations and properties of self-accelerating TCs. Here we first show that $\overrightarrow{C}<\infty$, while we later even obtain $\overrightarrow{C}=0$ when calculating the asymptotic profiles. $\overrightarrow{C}<\infty$ can be shown by presuming $\overrightarrow{C}=\infty$ and arriving at a contradiction. From $\overrightarrow{C}=\infty$ follows that
\begin{eqnarray}
 \overrightarrow{d\psi/dx}=\overrightarrow{E_s}v_s-\overrightarrow{r_o}v_s\overrightarrow{C}=-\infty, \label{help4}
\end{eqnarray}
where we used $0<\overrightarrow{E_s}<\infty$, $0<v_s<\infty$, $\overrightarrow{r_o}>0$, and Eq.~(\ref{Rpsihilf4}). Eq.~(\ref{help4}) is a contradiction to $\overrightarrow{\psi}=\infty$. Hence, $\overrightarrow{C}<\infty$.

\subsubsection{Showing $\overrightarrow{h}=\infty$}
$\overrightarrow{h}=\infty$ can be shown by presuming $\overrightarrow{h}=h_c$, where $h_c$ is a finite length, and arriving at a contradiction. Inserting this presumption in the limit $x\rightarrow\infty$ of Eq.~(\ref{conc}) yields
\begin{eqnarray}
 \overrightarrow{C}=\frac{1}{h_c}\overrightarrow{\psi/U}=\frac{1}{h_c\tilde g}\overrightarrow{RiU^2}=\infty,
\end{eqnarray}
where we used $\overrightarrow{\tilde g}<\infty$, $\overrightarrow{Ri}>0$, and $\overrightarrow{U}=\infty$. This is a contradiction to $\overrightarrow{C}<\infty$. Hence, $\overrightarrow{h}=\infty$.

\subsubsection{Showing $\overrightarrow{Ri}<\infty$ and $\overrightarrow{e_w}>0$}
$\overrightarrow{Ri}<\infty$ can be shown by presuming $\overrightarrow{Ri}=\infty$ and arriving at a contradiction. Inserting this presumption in the limit $x\rightarrow\infty$ of Eqs.~(\ref{massmixture}) and (\ref{masssediment}), respectively, yields
\begin{eqnarray}
 \overrightarrow{dh/dx}&=&S-\frac{1}{2}\overrightarrow{R_\psi}, \label{help1} \\
 \overrightarrow{\frac{h}{U}\frac{dU}{dx}}&=&-S+\frac{1}{2}\overrightarrow{R_\psi}=-\overrightarrow{dh/dx}\Rightarrow, \label{help2} \\
 \overrightarrow{\frac{h}{U}\frac{dU}{dx}}+\overrightarrow{dh/dx}&=&0, \label{help5}
\end{eqnarray}
where we used $S<\infty$, $C_D<\infty$, Eq.~(\ref{bedshearTEM}), and $e_w\xrightarrow{Ri\rightarrow\infty}0$. Due to $\overrightarrow{h}=\infty$ and $\overrightarrow{U}=\infty$, $\overrightarrow{dh/dx}\geq0$ and $\overrightarrow{\frac{h}{U}\frac{dU}{dx}}\geq0$ must be fulfilled. It then follows from Eq.~(\ref{help5}) that
\begin{eqnarray}
 \overrightarrow{dh/dx}=\overrightarrow{\frac{h}{U}\frac{dU}{dx}}=0 \label{help6}
\end{eqnarray}
and thus $\overrightarrow{R_\psi}=2S$ due to Eq.~(\ref{help1}). Since $0<S<\infty$, this allows us to calculate
\begin{eqnarray}
 \overrightarrow{d\psi/dx}=\overrightarrow{R_\psi\psi/h}=\overrightarrow{R_\psi}\,\overrightarrow{\psi/h}=\overrightarrow{R_\psi}\frac{1}{\overrightarrow{h/\psi}}=\overrightarrow{E_s}v_s, \label{help9}
\end{eqnarray}
where we used Eqs.~(\ref{momentummixture}) and (\ref{Rpsihilf3}). Using this and l'Hospital's rule \citep{Chatterjee2012}, we then also obtain from Eq.~(\ref{Rpsihilf3})
\begin{eqnarray}
 0<2S=\overrightarrow{R_\psi}=\overrightarrow{E_s}v_s\overrightarrow{h/\psi}=\overrightarrow{E_s}v_s\overrightarrow{\left(\frac{dh/dx}{d\psi/dx}\right)}=\overrightarrow{dh/dx}, \label{contra}
\end{eqnarray}
where we used $v_s>0$ and $\overrightarrow{E_s}>0$. Eq.~(\ref{contra}) is a contradiction to Eq.~(\ref{help6}). Hence, $\overrightarrow{Ri}<\infty$ and thus $\overrightarrow{e_w}>0$ (water is entrained if flow turbulence is present).

\subsubsection{Showing $\overrightarrow{R_\psi}<\infty$}
$\overrightarrow{R_\psi}<\infty$ can be shown by presuming $\overrightarrow{R_\psi}=\infty$ and arriving at a contradiction. Inserting this presumption in the limit $x\rightarrow\infty$ of Eqs.~(\ref{massmixture}) and (\ref{masssediment}) using Eq.~(\ref{bedshearTEM}) yields.
\begin{eqnarray}
 \overrightarrow{dh/dx}=\overrightarrow{\mathrm{sgn}(1-Ri)}\,\infty, \label{help7} \\
 \overrightarrow{\frac{h}{U}\frac{dU}{dx}}=-\overrightarrow{\mathrm{sgn}(1-Ri)}\,\infty \label{help8},
\end{eqnarray}
where we used $\overrightarrow{Ri}<\infty$ and $\mathrm{sgn}$ denotes the signum function. Eqs.~(\ref{help7}) and (\ref{help8}) mean that either $\overrightarrow{dh/dx}$ or $\overrightarrow{\frac{h}{U}\frac{dU}{dx}}$ will become $-\infty$ depending on the limit of the sign of $1-Ri$. However, a negative value of $\overrightarrow{dh/dx}$ or $\overrightarrow{\frac{h}{U}\frac{dU}{dx}}$ is a contradiction to $\overrightarrow{U}=\overrightarrow{h}=\infty$. Hence, $\overrightarrow{R_\psi}<\infty$.

\subsubsection{Asymptotic behaviors of $U$, $\psi$, and $h$}
The asymptotic behaviors of $U$, $\psi$, and $h$ can be calculated through computing $\overrightarrow{R_\psi}$ and $\overrightarrow{R_i}$ using l'Hospital's rule \citep{Chatterjee2012}. First, if $\overrightarrow{R_\psi}>0$, using $\overrightarrow{R_\psi}<\infty$, we can use the same arguments which we used prior to Eq.~(\ref{contra}) and calculate $\overrightarrow{R_\psi}$ by
\begin{eqnarray}
 \overrightarrow{R_\psi}=\overrightarrow{dh/dx}, \label{dhdx}
\end{eqnarray}
If $\overrightarrow{R_\psi}=0$, we cannot separate $\overrightarrow{R_\psi(\psi/h)}=\overrightarrow{R_\psi}\,\overrightarrow{\psi/h}$, as we did in Eq.~(\ref{help9}). However, in this case, we can calculate
\begin{eqnarray}
 0&=&\overrightarrow{R_\psi}=\overrightarrow{E_s}v_s\overrightarrow{h/\psi}=\overrightarrow{E_s}v_s\overrightarrow{\left(\frac{dh/dx}{d\psi/dx}\right)} \nonumber \\
 &=&\overrightarrow{E_s}v_s\overrightarrow{\left(\frac{dh/dx}{E_sv_s-r_o\,v_sC}\right)}, \label{dhdx2}
\end{eqnarray}
from Eqs.~(\ref{Rpsihilf3}) and (\ref{Rpsihilf4}), where we used $v_s>0$ and $\overrightarrow{E_s}>0$. The only way in which the right hand side of Eq.~(\ref{dhdx2}) vanishes is through $\overrightarrow{dh/dx}=0$ since $0\leq\overrightarrow{C}<\infty$. Hence Eq.~(\ref{dhdx}) is also fulfilled if $\overrightarrow{R_\psi}=0$.

\begin{figure}
 \begin{center}
  \includegraphics[width=1.0\columnwidth]{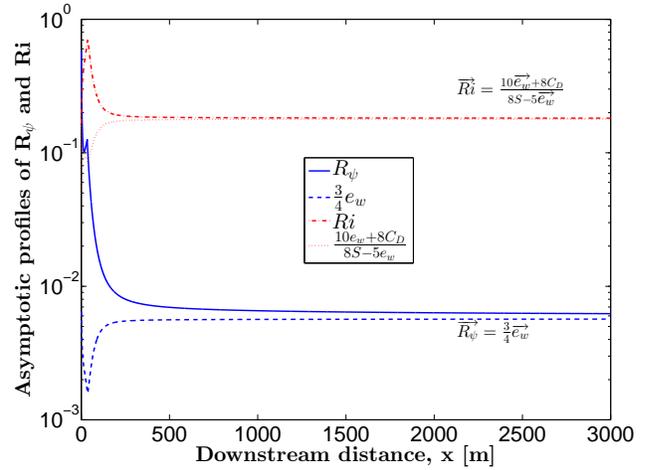}
 \end{center}
 \caption{Comparison between the downstream evolutions of $R_\psi$ and $Ri$ computed with the steady TEM using the parameter values and ignition values specified in F85 (see Table~\ref{parametervalues}) and their analytically derived asymptotic profiles (Eqs.~(\ref{Rpsi}) and (\ref{Ri2})).}
 \label{RpsiRi}
\end{figure}

\begin{figure}
 \begin{center}
  \includegraphics[width=1.0\columnwidth]{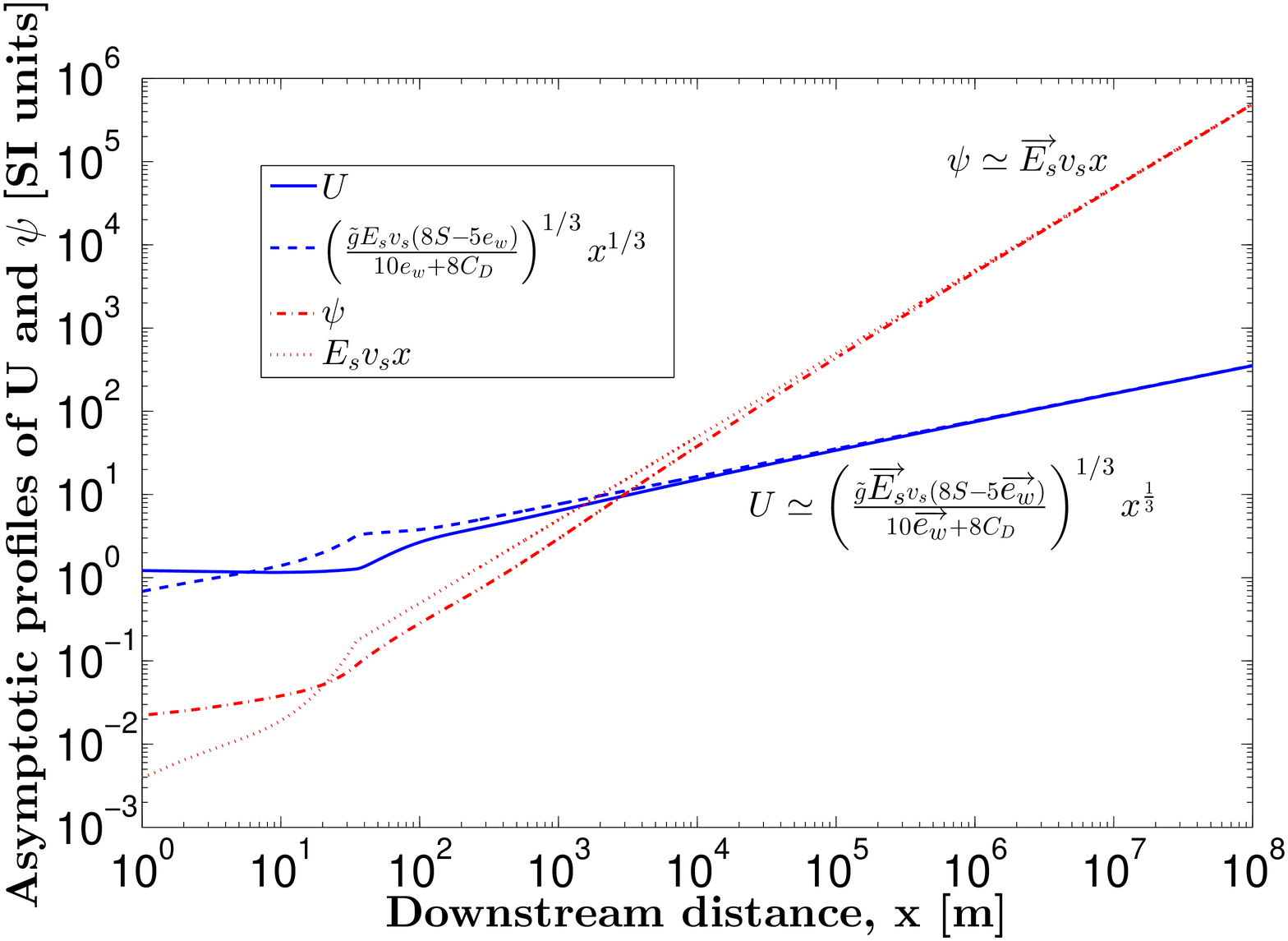}
 \end{center}
 \caption{Comparison between the downstream evolutions of $U$ and $\psi$ computed with the steady TEM using the parameter values and ignition values specified in F85 (see Table~\ref{parametervalues}) and their analytically derived asymptotic profiles (Eqs.~(\ref{Uasym}) and (\ref{psi2})).}
 \label{Upsi}
\end{figure}

\begin{figure}
 \begin{center}
  \includegraphics[width=1.0\columnwidth]{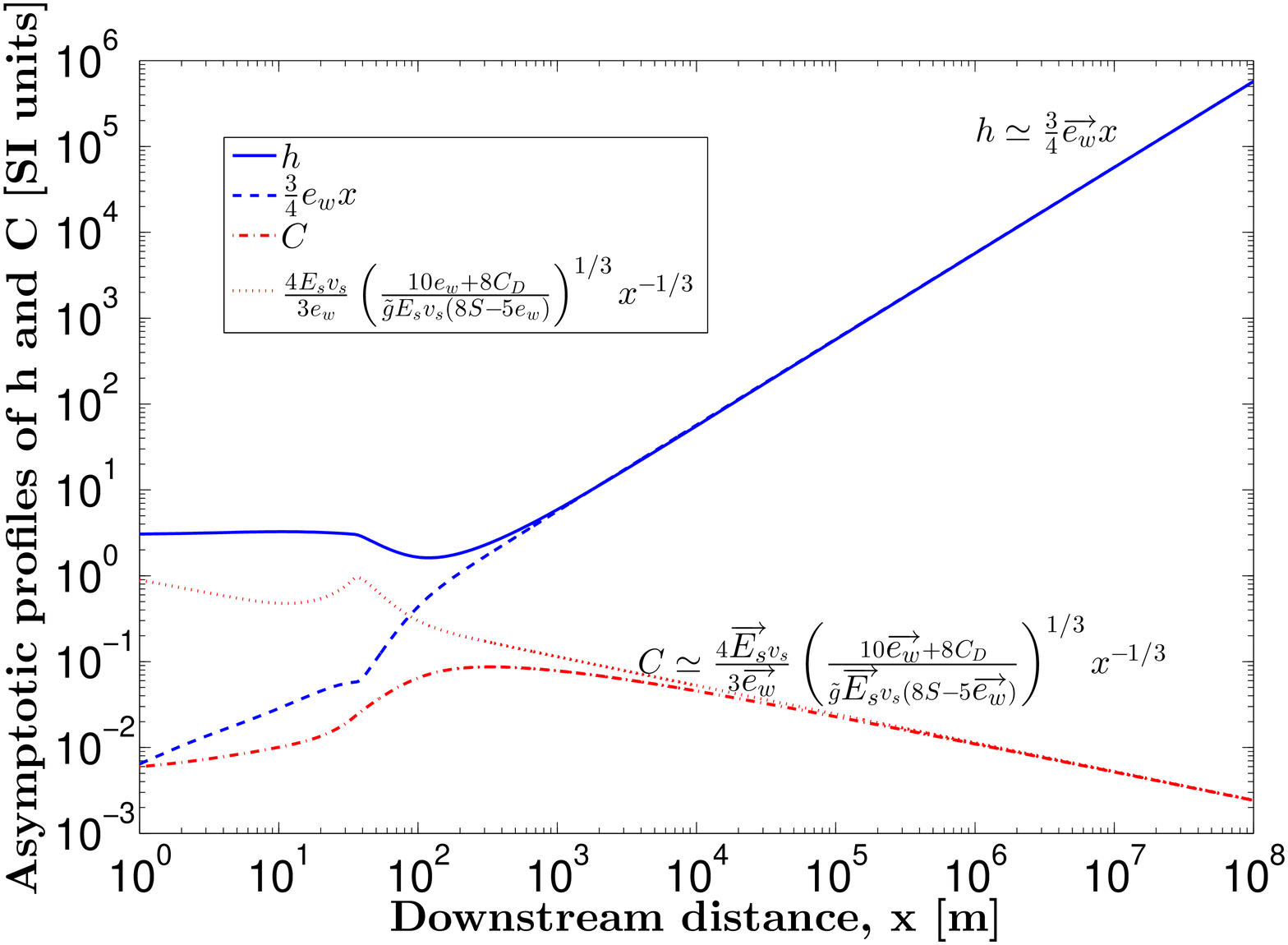}
 \end{center}
 \caption{Comparison between the downstream evolutions of $h$ and $C$ computed with the steady TEM using the parameter values and ignition values specified in F85 (see Table~\ref{parametervalues}) and their analytically derived asymptotic profiles (Eqs.~(\ref{hasym}) and (\ref{Casym})).}
 \label{hC}
\end{figure}

Now we calculate $\overrightarrow{Ri}$ in a similar manner and rearrange for $\frac{h}{U}\frac{dU}{dx}$ using $0<\overrightarrow{Ri}<\infty$. This yields
\begin{eqnarray}
 &&\overrightarrow{R_i}=\tilde g\overrightarrow{\psi/U^3}=\tilde g\overrightarrow{\frac{d\psi/dx}{3U^2dU/dx}}=\frac{1}{3}\tilde g\overrightarrow{\left(\psi/U^3\frac{R_\psi}{\frac{h}{U}dU/dx}\right)}\Rightarrow \nonumber \\
 &&\overrightarrow{\frac{h}{U}dU/dx}=\frac{\overrightarrow{R_\psi}}{3}, \label{dUdx}
\end{eqnarray}
where we used $0<\tilde g<\infty$ and Eq.~(\ref{Rpsihilf1}). From $\frac{dh}{dx}+\frac{h}{U}\frac{dU}{dx}=e_w$ (see Eqs.~(\ref{massmixture}) and (\ref{masssediment})) and Eqs.~(\ref{bedshearTEM}), (\ref{masssediment}), (\ref{dhdx}), and (\ref{dUdx}) then further follows
\begin{eqnarray}
 \overrightarrow{R_\psi}&=&0.75\overrightarrow{e_w}, \label{Rpsi} \\
 \overrightarrow{Ri}&=&\frac{10\overrightarrow{e_w}+8C_D}{8S-5\overrightarrow{e_w}}, \label{Ri2}
\end{eqnarray}
where we used $0<\overrightarrow{Ri}<\infty$, which implies $S>\frac{5}{8}\overrightarrow{e_w}$, and thus self-accelerating TCs do not exist if $S\leq\frac{5}{8}\overrightarrow{e_w}$. Finally, from Eqs.~(\ref{conc}), (\ref{Rpsihilf4}), (\ref{dhdx}), (\ref{dUdx}), (\ref{Rpsi}), and (\ref{Ri2}), one obtains that $U$, $\psi$, and $h$ must follow the following asymptotic behaviors,
\begin{eqnarray}
 \psi_\infty&=&\overrightarrow{E_s}v_sx, \label{psi2} \\
 h_\infty&=&\overrightarrow{R_\psi}x=0.75\overrightarrow{e_w}x, \label{hasym} \\
 U_\infty&=&\left(\frac{\tilde g\overrightarrow{E_s}v_s}{\overrightarrow{Ri}}\right)^{1/3}x^{\frac{1}{3}}=\left(\frac{\tilde g\overrightarrow{E_s}v_s(8S-5\overrightarrow{e_w})}{10\overrightarrow{e_w}+8C_D}\right)^{1/3}x^{\frac{1}{3}}, \nonumber \\
 && \label{Uasym}
\end{eqnarray}
where the subscript '$\infty$' indicates the asymptotic behavior. We note that Eq.~(\ref{psi2}) is consistent with Eq.~(\ref{Rpsihilf4}) since
\begin{eqnarray}
 C_\infty=\frac{\psi_\infty}{h_\infty U_\infty}=\frac{4\overrightarrow{E_s}v_s}{3\overrightarrow{e_w}}\left(\frac{10\overrightarrow{e_w}+8C_D}{\tilde g\overrightarrow{E_s}v_s(8S-5\overrightarrow{e_w})}\right)^{1/3}x^{-\frac{1}{3}}, \label{Casym}
\end{eqnarray}
which vanishes in the limit $x\rightarrow\infty$. We further note that the derived asymptotic profiles (Eqs.~(\ref{Rpsi}-\ref{Casym})) are in agreement with numerical steady TEM simulations of self-accelerating TCs, as can be seen in Figs.~\ref{RpsiRi}-\ref{hC}, which supports the correctness of our derivations. In these figures, we compare the downstream evolutions of $R_\psi$, $Ri$, $\psi$, $h$, $U$, and $C$ computed with the steady TEM using the parameter values and ignition values specified in F85 (see Table~\ref{parametervalues} in Section~\ref{simulations}) and their analytically derived asymptotic profiles (Eqs.~(\ref{Rpsi}-\ref{Casym})).

\subsection{Existence criteria for self-accelerating TCs} \label{existence}
In this part of the proof, we show that $\overrightarrow{Ri}<1$ is a necessary and sufficient condition for the existence of self-accelerating TCs using the results of Section~\ref{bounded}. Since the densimetric Froude number for gravity currents is defined as $Fr=1/\sqrt{Ri}$ \citep{KosticParker2006}, $Ri<1$ corresponds to supercritical flow. Our strategy to show $\overrightarrow{Ri}<1$ is as follows. First, we slightly modify the definition of $R_\psi$ in Eq.~(\ref{momentummixture}) in a manner that ensures that the asymptotic self-accelerating solution of Eqs.~(\ref{massmixture}-\ref{momentummixture}), given by $(h_\infty,U_\infty,\psi_\infty)$  (see Eqs.~(\ref{psi2}-\ref{Uasym}), becomes an exact solution of the modified problem. Since the original problem becomes arbitrarily close to the modified problem for self-accelerating TCs sufficiently far downstream, it can be considered as a small perturbation of the modified problem. Hence, a self-accelerating solution of the original problem exists if and only if $(h_\infty,U_\infty,\psi_\infty)$ is stable against small perturbations of the modified problem, which we show to be equivalent to $\overrightarrow{Ri}<1$.

\subsubsection{The modified problem}
We define the modified value of $R_\psi$, indicated by a tilde, as
\begin{eqnarray}
 \tilde R_\psi=\overrightarrow{E_s}v_sh/\psi. \label{Rpsimod}
\end{eqnarray}
This definition ensures that $\tilde R_\psi$ becomes arbitrarily close to $R_\psi$ (see Eq.~(\ref{Rpsihilf1})) for self-accelerating TCs because $E_s$ becomes arbitrarily close to $0<\overrightarrow{E_s}<\infty$, while $v_sr_o/U$ becomes arbitrarily small due to $\overrightarrow{U}=\infty$, $v_s<\infty$, and $\overrightarrow{r_o}<\infty$. Using Eqs.~(\ref{bedshearTEM}) and (\ref{Rpsimod}), Eqs.~(\ref{massmixture}-\ref{momentummixture}) are redefined as
\begin{eqnarray}
 \frac{dh}{dx}&=&\frac{-RiS+e_w(2-0.5Ri)+C_D+\frac{1}{2}Ri\tilde R_\psi}{1-Ri}, \label{massmixturemod} \\
 \frac{dU}{dx}&=&\frac{U}{h}\frac{RiS-e_w(1+0.5Ri)-C_D-\frac{1}{2}Ri \tilde R_\psi}{1-Ri}, \label{masssedimentmod} \\
 \frac{d\psi}{dx}&=&\overrightarrow{E_s}v_s, \label{momentummixturemod}
\end{eqnarray}

\subsubsection{Stability}
It can be easily verified that $(h_\infty,U_\infty,\psi_\infty)$, given by Eqs.~(\ref{psi2}-\ref{Uasym}), is an exact solution of Eqs.~(\ref{massmixturemod}-\ref{momentummixturemod}). We now determine the eigenvalues ($\lambda$) of the Jacobi matrix ($\partial(dh/dx,dU/dx,d\psi/dx)/\partial(h,U,\psi)$) evaluated at $(h_\infty,U_\infty,\psi_\infty)$, reading
\begin{eqnarray}
 \left|\lambda\mathbf{I}-\frac{\partial(dh/dx,dU/dx,d\psi/dx)}{\partial(h,U,\psi)}\right|(h_\infty,U_\infty,\psi_\infty)=0, \label{Eigenvalueeq}
\end{eqnarray}
where $\mathbf{I}$ is the identity matrix, and $|\cdot|$ denotes the determinant. If the real parts of all (some) eigenvalues are negative (positive), small perturbations from the solution $(h_\infty,U_\infty,\psi_\infty)$ decline (grow) with $x$, which means $(h_\infty,U_\infty,\psi_\infty)$ is stable (unstable). However, if the real parts of some of these eigenvalues vanish, one has to take a closer look. Indeed, one of the eigenvalues we obtain from Eq.~(\ref{Eigenvalueeq}) vanishes ($\lambda_1=0$) because $\partial(d\psi/dx)/\partial(h,U,\psi)=0$ (see Eq.~(\ref{momentummixturemod})). However, this does not make the solution unstable because $\partial(d\psi/dx)/\partial(h,U,\psi)=0$ ensures that any deviation from $\psi_\infty$ remains constant with $x$. The other two eigenvalues ($\lambda_{2/3}$) we obtain from Eq.~(\ref{Eigenvalueeq}) read
\begin{eqnarray}
 &&\lambda_{2/3}=\frac{-A\pm\sqrt{A^2-B}}{16h(1-\overrightarrow{Ri})}, \\
 &&A=24C_D+\overrightarrow{e_w}(28-\overrightarrow{Ri})-12\overrightarrow{e_w^\prime}\overrightarrow{Ri}(2+\overrightarrow{Ri}), \nonumber \\
 &&B=24\overrightarrow{e_w}(1-\overrightarrow{Ri})(8C_D-8\overrightarrow{e_w^\prime}\overrightarrow{Ri}(2+\overrightarrow{Ri})+\overrightarrow{e_w}(10+\overrightarrow{Ri})), \nonumber
\end{eqnarray}
where we used $e_w^\prime=de_w/dRi\leq0$, the definitions of $Ri$ and $\tilde R_\psi$, and Eqs.~(\ref{Rpsi}-\ref{Uasym}). One can see that, if $\overrightarrow{Ri}<1$, the real parts of both $\lambda_2$ and $\lambda_3$ are negative and the solution $(h_\infty,U_\infty,\psi_\infty)$ thus stable. However, if $\overrightarrow{Ri}>1$, the eigenvalue $\lambda_3=(-A-\sqrt{A^2-B})/(16h(1-\overrightarrow{Ri}))$ is always positive due to $B<0$ and the solution $(h_\infty,U_\infty,\psi_\infty)$ thus unstable. Furthermore, if $\overrightarrow{Ri}=1$, the solution is still unstable because small perturbations for which $Ri$ approaches unity from above still result in a positive eigenvalue $\lambda_3$. Hence,
\begin{eqnarray}
 \overrightarrow{Ri}<1 \label{Ricriteria}
\end{eqnarray}
is a criterion for the stability and thus existence of self-accelerating TCs simulated with the steady TEM. This criterion leads to another criterion for $S$, reading
\begin{eqnarray}
 S>\frac{15}{8}\overrightarrow{e_w}+C_D\geq\frac{15}{8}e_w|_{Ri=1}+C_D, \label{Scriteria}
\end{eqnarray}
where we used $de_w/dRi\leq0$ and Eq.~(\ref{Ri2}).

\subsection{Sufficient conditions for $C_D$ or alternatively $S$} \label{necessary}
In this part of the proof, we presume $\overrightarrow{\Delta E}\leq0$ and will arrive at necessary conditions for $C_D$ and $S$ using the results of Sections~\ref{bounded} and \ref{existence}. Hence, the opposite conditions for $C_D$ or alternatively $S$ are sufficient for $\overrightarrow\Delta E>0$. First, we show that $\overrightarrow{\Delta E}<0$ is never fulfilled since it leads to a contradiction: In fact, $\overrightarrow{\Delta E}<0$ necessarily implies that $\overrightarrow{dk/dx}<0$ (see Eq.~\ref{energymixture}) and thus $\overrightarrow{k}=-\infty$ (since $\overrightarrow{k}=k_c$, where $k_c$ is an arbitrary finite value, would imply $\overrightarrow{dk/dx}=0$), which is a contradiction since the Heaviside function $\Theta(k)$ in Eq.~(\ref{energymixture}) ensures that $\Delta E=dk/dx=0$ if $k=0$. Hence, we only have to consider the case $\overrightarrow{\Delta E}=0$. In this case, we obtain $\overrightarrow{k/U^2}=0$ because, if $\overrightarrow{k}=\infty$, 
\begin{eqnarray}
 && \overrightarrow{\left(\frac{k}{U^2}\right)}=\frac{1}{2}\overrightarrow{\left(\frac{dk/dx}{UdU/dx}\right)}=\frac{1}{2}\overrightarrow{\left(\frac{\frac{h}{U^2}\frac{dk}{dx}}{\frac{h}{U}\frac{dU}{dx}}\right)}=2\overrightarrow{\left(\frac{\Delta E}{e_w}\right)}=0, \nonumber \\
 && \label{kU}
\end{eqnarray}
where we used $\overrightarrow{e_w}>0$, Eqs.~(\ref{energymixture}), (\ref{dUdx}), and (\ref{Rpsi}) and l'Hospital's rule \citep{Chatterjee2012}, and if $\overrightarrow{k}<\infty$, $\overrightarrow{k/U^2}=0$ anyways. Now we perform the limit $x\rightarrow\infty$ on Eq.~(\ref{energymixture}) using $\overrightarrow{U}=\infty$, Eq.~(\ref{bedshearTEM}), $\overrightarrow{\Delta E}=0$, and $\overrightarrow{RiR_\psi}=\overrightarrow{Ri}\,\overrightarrow{R_\psi}$ due to $\overrightarrow{Ri}<\infty$ and $\overrightarrow{R_\psi}<\infty$, yielding
\begin{eqnarray}
 0=\overrightarrow{\Delta E}=\max\left(0,\frac{1}{2}\overrightarrow{e_w}(1-\overrightarrow{Ri})+C_D-\frac{1}{2}\overrightarrow{Ri}\,\overrightarrow{R_\psi}\right), \label{eqstep1}
\end{eqnarray}
where we further used $\overrightarrow{e_w}<\infty$, $\overrightarrow{Ri}<\infty$, and $v_s<\infty$ and thus $\overrightarrow{Riv_s/U}=0$, and that $\overrightarrow{ke_w/U^2}=0$ and $\overrightarrow{\beta k^{3/2}/U^3}=\overrightarrow{\beta (k/U^2)^{3/2}}=0$ due to $\overrightarrow{e_w}<\infty$, $\overrightarrow{\beta}<\infty$, and Eq.~(\ref{kU}). The maximum function in Eq.~(\ref{eqstep1}) occurs because, if $0.5\overrightarrow{e_w}(1-\overrightarrow{Ri})+C_D-0.5\overrightarrow{Ri}\,\overrightarrow{R_\psi}$ is negative, $k$ will continue to decrease until it vanishes, in which case $\Theta(k)$ and thus $\Delta E$ calculated by Eq.~(\ref{energymixture}) also vanish. Eq.~(\ref{eqstep1}) provides a lower limit for $\frac{1}{2}\overrightarrow{Ri}\,\overrightarrow{R_\psi}$, reading
\begin{eqnarray}
 &&0.5\overrightarrow{e_w}(1-\overrightarrow{Ri})+C_D\leq0.5\overrightarrow{Ri}\,\overrightarrow{R_\psi}. \label{eqstep2}
\end{eqnarray}
Now we insert Eq.~(\ref{Rpsi}) into Eq.~(\ref{eqstep2}) and rearrange for $\overrightarrow{Ri}$, yielding
\begin{eqnarray}
 \overrightarrow{Ri}&\geq&\frac{4}{7}\left(\frac{2C_D}{\overrightarrow{e_w}}+1\right)>\frac{4}{7}, \label{Ri} 
\end{eqnarray}
where we used $C_D>0$ and $0<\overrightarrow{e_w}<\infty$.

\subsubsection{A sufficient condition for $C_D$}
Rearranging Eq.~(\ref{Ri}) for $C_D$ and using $de_w/dRi\leq0$ and Eq.~(\ref{Ricriteria}) yields
\begin{eqnarray}
 C_D<\frac{3}{8}\overrightarrow{e_w}\leq\frac{3}{8}e_w|_{Ri=\frac{4}{7}}=C_{D\mathrm{min}}. \label{Cd}
\end{eqnarray}
Eq.~(\ref{Cd}) is a condition which must be fulfilled if $\overrightarrow{\Delta E}\leq0$. This means, if $C_D\geq C_{D\mathrm{min}}$, self-accelerating TCs simulated with the steady TEM always fulfill $\overrightarrow{\Delta E}>0$.

Finally, rearranging Eq.~(\ref{Ri2}) for $S$ and using Eq.~(\ref{Ri}) yields
\begin{eqnarray}
 &&S=\frac{5}{8}\overrightarrow{e_w}+\frac{\frac{5}{4}\overrightarrow{e_w}+C_D}{\overrightarrow{Ri}}\leq\frac{5}{8}\overrightarrow{e_w}+\frac{7}{4}\overrightarrow{e_w}\frac{\frac{5}{4}\overrightarrow{e_w}+C_D}{\overrightarrow{e_w}+2C_D} \nonumber \\
 &&\leq\frac{45}{16}e_w|_{Ri=\frac{4}{7}}=S_\mathrm{min}, \label{S}
\end{eqnarray}
where we used $de_w/dRi\leq0$ and that
\begin{eqnarray}
 \frac{\partial}{\partial C_D}\left(\frac{\frac{5}{4}\overrightarrow{e_w}+C_D}{\overrightarrow{e_w}+2C_D}\right)=-\frac{\frac{3}{2}\overrightarrow{e_w}}{(\overrightarrow{e_w}+2C_D)^2}<0
\end{eqnarray}
and thus
\begin{eqnarray}
 \frac{\frac{5}{4}\overrightarrow{e_w}+C_D}{\overrightarrow{e_w}+2C_D}<\frac{\frac{5}{4}\overrightarrow{e_w}+0}{\overrightarrow{e_w}+0}=\frac{5}{4}.
\end{eqnarray}
Eq.~(\ref{S}) is a condition which must be fulfilled if $\overrightarrow{\Delta E}\leq0$. This means, if $S\geq S_\mathrm{min}$, self-accelerating TCs simulated with the steady TEM always fulfill $\overrightarrow{\Delta E}>0$.

\subsection{Are $C_D\geq C_{D\mathrm{min}}$ or $S\geq S_\mathrm{min}$?} \label{relevantcases}
Now we argue that at least one of the conditions $C_D\geq C_{D\mathrm{min}}$ and $S\geq S_\mathrm{min}$ is virtually always fulfilled. According to Eqs.~(\ref{Cd}) and (\ref{S}), the values of $C_{D\mathrm{min}}$ and $S_\mathrm{min}$ depends on the empirical relationship used to calculate $e_w$. Standard relationships for $e_w$ are, for instance, Eq.~(\ref{ew}), for which $C_{D\mathrm{min}}=0.00097$ and $S_\mathrm{min}=0.0073$, and $e_w=0.075/\sqrt{1+718Ri^{2.4}}$ \citep{Parkeretal1987}, for which $C_{D\mathrm{min}}=0.00205$ and $S_\mathrm{min}=0.0154$. However, recent state of the art simulations of suspended sediment transport \citep{Schmeeckle2014} (see Section~\ref{comparison}) as well as the majority of experimental data \citep{BradfordKatopodes1999} indicate that realistic values of $C_D$ are significantly larger. Hence, in realistic simulations self-accelerating TCs always fulfill $\overrightarrow{\Delta E}>0$. Even if an unrealistically small value for $C_D$ is used, one would still have to consider bed slopes which are much smaller than those one typically uses for TCs. For instance, F85 used $C_D=0.004$ and $S=0.08$. Each of these values alone is actually sufficient to ensure that their simulated self-accelerating TCs were physically realistic ($\overrightarrow{\Delta E}>0$). The fact that F85 reported $\overrightarrow{\Delta E}<0$ thus means that there was most likely an error in the numerical computations. Indeed, we show in the following that, depending on the ignition values, numerical simulations using the same parameter values and empirical relations as reported by F85 and P86 result either in self-accelerating TCs which fulfill $\overrightarrow{\Delta E}>0$ or in TCs which fulfill $\overrightarrow{U}<\infty$ and thus are not self-accelerating.

\section{Numerical simulations using the steady TEM} \label{simulations}
In this section, we first report simulations of self-accelerating TCs using the steady TEM and exactly the same empirical relations (Eqs.~(\ref{ew}-\ref{Es})), physical parameters and ignition values (see Table~\ref{parametervalues}) as F85 and P86 in Section~\ref{originalTEM}.
\begin{table}
\caption{Parameter values and ignition values specified in F85 and P86. The values in brackets correspond to the modified ignition values which we used to obtain self-accelerating TCs for the parameter values specified in P86.}
\centering
\begin{tabular}{|c|c|c|}
\hline
  & \citet{Fukushimaetal1985} & \citet{Parkeretal1986}  \\
\hline
  $S$  & 0.08 & 0.05 \\
  $D_s$ $\mathrm{[mm]}$  & $0.15$ & $0.1$ \\
  $v_s$ $\mathrm{[m/s]}$  & $0.0165$ & $0.0084$ \\
  $\nu$ $\mathrm{[m^2/s]}$  & $10^{-6}$ & $10^{-6}$ \\
  $\rho_s$ $\mathrm{[kg/m^3]}$  & $2650$ & $2650$ \\
  $\rho_w$ $\mathrm{[kg/m^3]}$  & $1000$ & $1000$ \\
  $C_D$ & $0.004$ & $0.004$ \\
  $h(0)$ $\mathrm{[m]}$  & $3$ & $2$ $(1)$ \\
  $U(0)$ $\mathrm{[m/s]}$  & $1.24$ & $0.652$ $(0.699)$ \\
  $\psi(0)$ $\mathrm{[m^2/s]}$  & $0.019$ & $0.0038$ $(0.0047)$ \\
\hline
\end{tabular}
\label{parametervalues}
\end{table}
Afterwards in Section~\ref{BC}, we present a new method to obtain ignition values, which even results in self-accelerating TCs if $Ri$ is very close to unity.

In all our simulations, the model equations were integrated using the Runge-Kutta method, and we confirmed that reducing the spatial integration step did not significantly change the results.

\subsection{Repetition of the original simulations by F85 and P86} \label{originalTEM}
\begin{figure}
 \begin{center}
  \includegraphics[width=1.0\columnwidth]{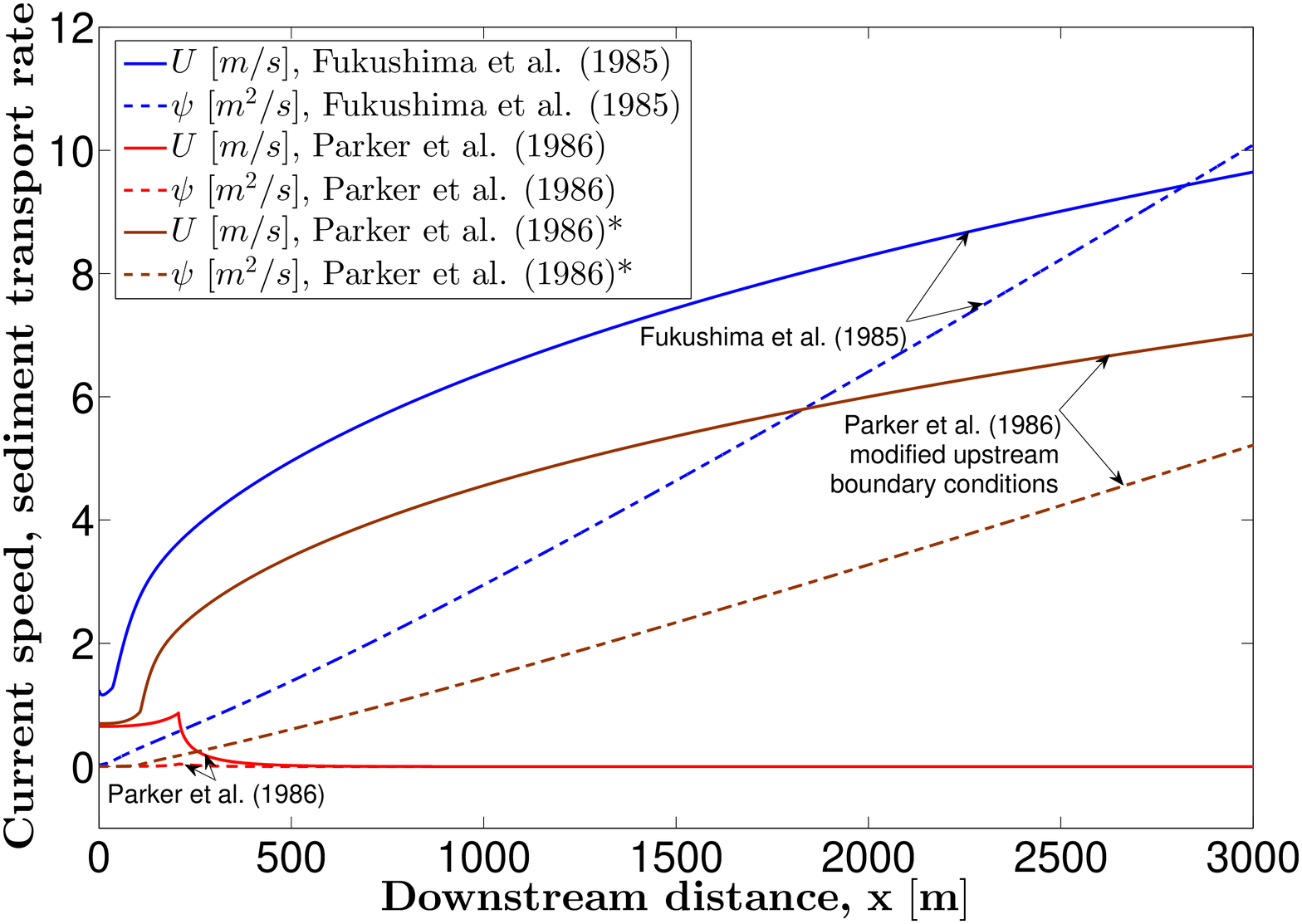}
 \end{center}
 \caption{TCs simulated with the steady TEM for the parameter values and ignition values specified in F85 and P86 (see Table~\ref{parametervalues}). The solid lines show the downstream profiles of $U$ and the dashed lines the downstream profiles of $\psi$. The star labels the case in which the ignition values used in P86 are modified to those given in brackets in Table~\ref{parametervalues}.}
 \label{TEMsimulations}
\end{figure}

Fig.~\ref{TEMsimulations} shows the downstream evolutions of $U$ (solid lines) and $\psi$ (dashed lines) for the TCs simulated with the steady TEM. As can be seen, while the parameter values and ignition values specified in F85 result in self-accelerating TCs, those specified in P86 do not since $U$ and $\psi$ are decreasing downstream for $x>212$m.
In order to avoid confusion, it is very important to realize here that P86 called this TC ``self-accelerating'' because of its accelerating behavior for $x<212$m, while it is \textit{not} self-accelerating according to our definition. Moreover, these authors found that the decrease of $U$ and $\psi$ for $x>212$m coincides with $\Delta E<0$. However, it is not surprising at all that decelerating TCs loose turbulent kinetic energy when moving downstream. In fact, also the steady FEM produces decelerating TCs if the initial conditions are chosen appropriately \citep{Parkeretal1986}, and these currents should generally also loose turbulent kinetic energy when moving downstream. There is no qualitative difference between the steady TEM and FEM in this regard. Hence, in order to disproof the claim of P86 that the steady TEM cannot produce physically realistic ($\overrightarrow{\Delta E}>0$) self-accelerating TCs, we only have to show that, for the same physical parameters ($S$, $D_s$, $v_s$, $\nu$), there are ignition values which result in self-accelerating TCs since these automatically fulfill $\overrightarrow{\Delta E}>0$ (see our proof in Section~\ref{Proof}). Indeed, by changing the ignition values (see the values in brackets in Table~\ref{parametervalues}), the parameter values specified in P86 result in self-accelerating TCs (the case labeled by a star in the legend of Fig.~\ref{TEMsimulations}). To be consistent, we used the same procedure as P86 (the method by \citet{Parker1982}) to obtain the modified ignition values, which is defined in the following. First, one chooses arbitrarily the upstream boundary condition $h(0)$. Then one defines a quasi-equilibrium state by $(dU/dx)(0)=(d\psi/dx)(0)=0$, but $(dh/dx)(0)\ne0$. Using this Eqs.~(\ref{masssediment}) and (\ref{momentummixture}) can be solved for $U(0)$ and $\psi(0)$. While P86 originally chose $h(0)=2\mathrm{m}$, from which they computed $U(0)=0.652\mathrm{m/s}$ and $\psi(0)=0.0038\mathrm{m^2/s}$, the modified value reads $h(0)=1\mathrm{m}$, from which one computes $U(0)=0.699\mathrm{m/s}$ and $\psi(0)=0.0047\mathrm{m^2/s}$ (see Table~\ref{parametervalues}).

Note that the fact that the original ignition values specified by P86 result in a decelerating TC shows that the method by \cite{Parker1982} to obtain the ignition values does not always work very well. The reason is that this method does not necessarily ensure that the ignition state is sufficiently close to the asymptotic solution ($h_\infty,\psi_\infty,U_\infty$). In Section~\ref{BC} we will provide an improved method to obtain the ignition values that even works in cases in which the method by \cite{Parker1982} does not at all, namely when $\overrightarrow{Ri}$ is close to unity.

\begin{figure}
 \begin{center}
  \includegraphics[width=1.0\columnwidth]{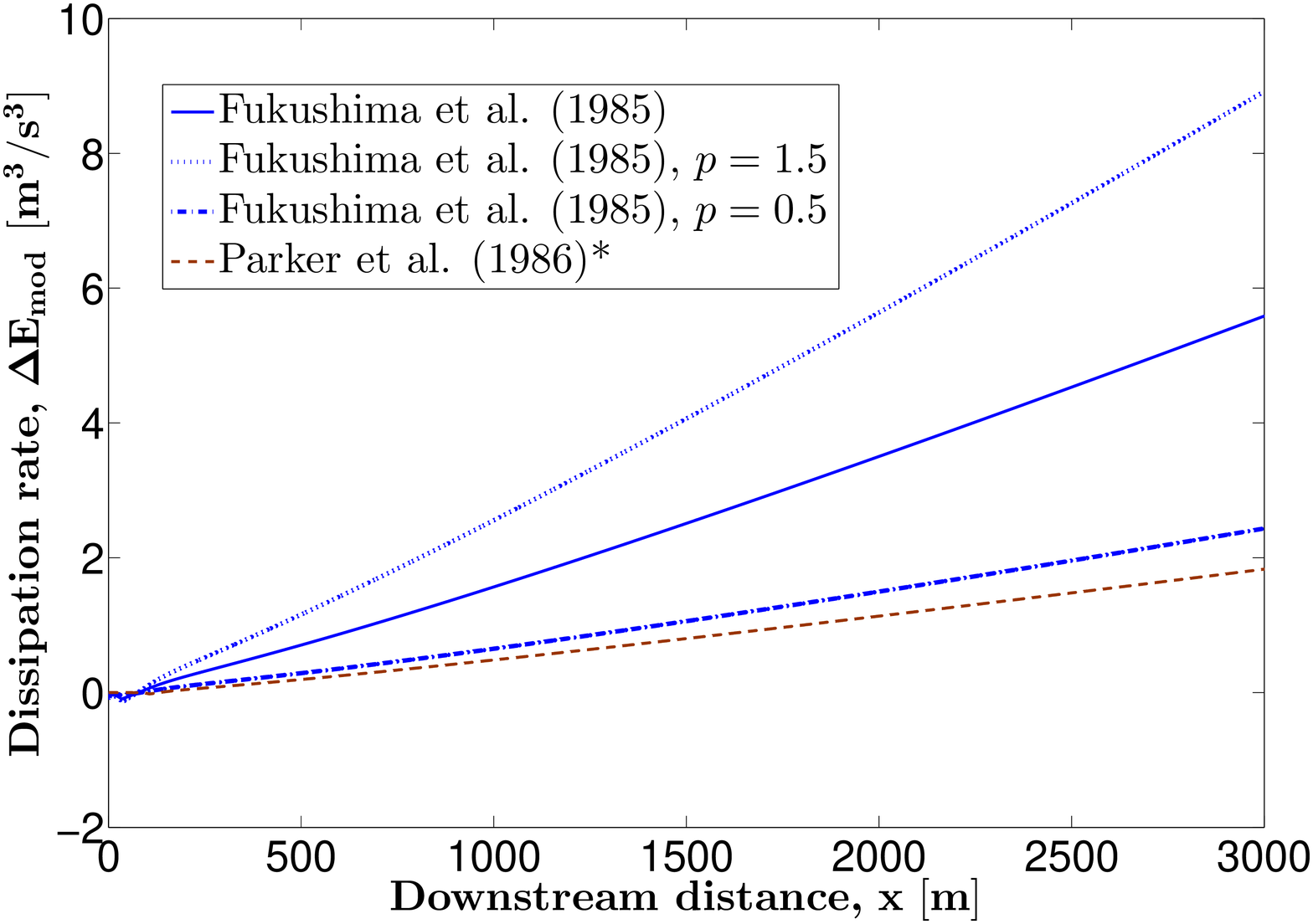}
 \end{center}
 \caption{TCs simulated with the steady TEM for the parameter values and ignition values specified in F85 and for the parameter values specified in P86 with the modified ignition values (see Table~\ref{parametervalues}), indicated by a star. The cases $p=1.5$ and $p=0.5$ refer to the use of a modified bed sediment erosion rate relation (Eq.~(\ref{Esmod})).}
 \label{deltaEmod}
\end{figure}

It remains to show that the self-accelerating TCs shown in Fig.~\ref{TEMsimulations} (i.e., the F85 current and the modified P86 current), but not the original P86 current (which is not self-accelerating, but decelerating), fulfill $\overrightarrow{\Delta E}>0$. In order to do so, we note that, if
\begin{eqnarray}
 \Delta E_{\mathrm{mod}}=U^3\left(\frac{1}{2}e_w(1-Ri)+C_D-Ri\frac{v_s}{U}-\frac{1}{2}RiR_\psi\right) \label{criteria}
\end{eqnarray}
is smaller than zero, then, due to Eqs.~(\ref{bedshearTEM}) and (\ref{energymixture}) and $e_w>0$, $\beta>0$, and $k>0$,
\begin{eqnarray}
 \Delta E=\frac{1}{U^3}\Delta E_{\mathrm{mod}}-\frac{ke_w}{U^2}-\frac{\beta k^{3/2}}{U^3}
\end{eqnarray}
is also smaller than zero. It follows that $\Delta E_{\mathrm{mod}}\geq0$ is a condition which must be fulfilled ever after a finite distance downstream in order for self-accelerating TCs to be physically realistic ($\overrightarrow{\Delta E}>0$). In fact, this criterion was used by F85 and P86 to determine whether self-accelerating TCs are physically realistic. Indeed, in accordance with our analytical proof in Section~\ref{Proof}, Fig.~\ref{deltaEmod} shows that $\Delta E_{\mathrm{mod}}\geq0$ is fulfilled.
It shows the downstream profiles of $\Delta E_{\mathrm{mod}}$ for cases in Table~\ref{parametervalues} that resulted in self-accelerating TCs. It can be seen that the qualitative behavior of $\Delta E_{\mathrm{mod}}$ is exactly opposite to the one described in F85. Instead of $\Delta E_{\mathrm{mod}}$ being positive in the beginning and more and more negative later on downstream, as described by F85, $\Delta E_{\mathrm{mod}}$ is negative in the beginning and more and more positive later on downstream. This strongly suggests an error in the early computations by F85.

Fig.~\ref{deltaEmod} also shows the downstream profiles of $\Delta E_{\mathrm{mod}}$ for the parameter values and ignition values specified in F85, but with a modified bed sediment erosion rate relation,
\begin{eqnarray}
  E_s=\left\{ 
  \begin{array}{l l}
    0.3p & Z\geq13.2 \\
    3p\times10^{-12}Z^{10}(1-5/Z) &  5<Z<13.2 \\
    0 & Z\leq5
  \end{array} \right\}, \label{Esmod}
\end{eqnarray}
where $p>0$ is a constant factor (the case $p=1$ is identical to Eq.~(\ref{Es})). It can be seen that the larger is the value of $p$ and thus $E_s$ the larger is $\Delta E_{\mathrm{mod}}$. This is again in contrast to the claim of F85 and P86 that their ostensible failure of the steady TEM is due to strong erosion of bed sediment. According to their argument, stronger erosion should lead to more energy spent in eroding and suspending bed sediment and thus to smaller values of $\Delta E_{\mathrm{mod}}$. However, one must also take into account that the eroded bed sediment has a potential energy which is converted into turbulent kinetic energy when moving downslope. This conversion would not take place if the sediment continued to rest at the top of the sediment bed. It can be seen in Fig.~\ref{deltaEmod} that this manner of additional production of turbulent kinetic energy more than compensates the additional energy loss due to the stronger erosion of bed sediment. Indeed, as we show in Section~\ref{explanation}, $\Delta E_{\mathrm{mod}}$ is proportional to $E_s$.

\subsection{Ignition of self-accelerating TCs} \label{BC}
In Section~\ref{existence}, we showed that self-accelerating TCs exist for $\overrightarrow{Ri}<1$ when simulated with the steady TEM. This might seem quite surprising because the condition $Ri>0.25$ is generally used to estimate whether the turbulent mixing becomes insufficient to overcome density layering in physical environments \citep{Prandle2009}. For this reason, we show here that numerical simulations with the steady TEM, indeed, result in self-accelerating TCs if the ignition values are chosen appropriately, even if $\overrightarrow{Ri}$ is very close to unity.

In order for $(h,U,\psi)$ to converge against the asymptotic solution $(h_\infty,U_\infty,\psi_\infty)$, the upstream boundary values of $Ri$ and $R_\psi$ should not deviate too strongly from their asymptotically constant values $\overrightarrow{Ri}$ and $\overrightarrow{R_\psi}$ since we only showed stability against sufficiently small deviations in Section~\ref{existence}. We thus propose to use the following ignition values
\begin{eqnarray}
 Ri(0)=\overrightarrow{Ri}&=&\frac{5a}{16S}+\frac{C_D}{2S}-\frac{b}{2} \nonumber \\
 &&+\sqrt{\left(\frac{5a}{16S}\frac{C_D}{2S}-\frac{b}{2}\right)^2+\frac{5a}{4S}+\frac{C_Db}{S}}, \label{Ribc} \\
 R_\psi(0)=\overrightarrow{R_\psi}&=&\frac{3a}{4(b+Ri(0))}, \label{Rpsibc}
\end{eqnarray}
where $a=0.00153$, $b=0.0204$, and the right hand sides are the solutions for $\overrightarrow{R_\psi}$ and $\overrightarrow{Ri}$ of Eqs.~(\ref{Rpsi}) and (\ref{Ri2}), respectively, when $e_w$ is calculated by Eq.~(\ref{ew}). For a given value of $h(0)$, Eqs.~(\ref{Ribc}) and (\ref{Rpsibc}) can be solved to obtain the upstream boundary values $U(0)$ and $\psi(0)$ using $Ri=\tilde g\psi/U^3$ and Eq.~(\ref{momentummixture}). Thereby larger values of $h(0)$ correspond to larger values of $U(0)$ and $\psi(0)$ and are thus relatively closer to the asymptotic solution. Hence, a sufficiently large value of $h(0)$ ensures that the ignition values calculated in the manner above always leads to self-accelerating TCs.

\begin{figure}
 \begin{center}
  \includegraphics[width=1.0\columnwidth]{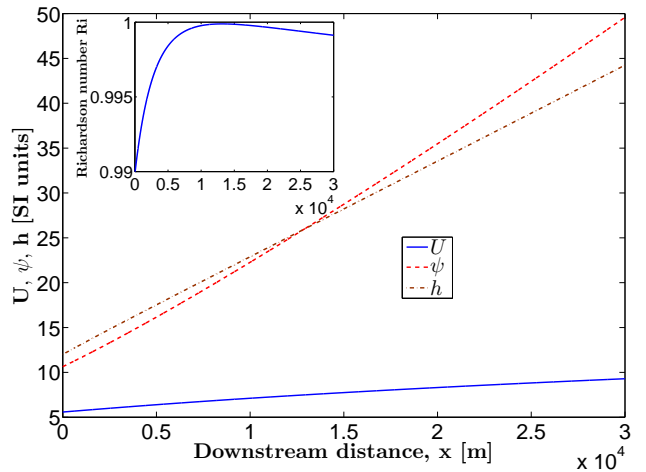}
 \end{center}
 \caption{Downstream evolutions of $U$, $\psi$, and $h$ (inset: $Ri$) computed with the steady TEM using the parameter values specified in P86 (see Table~\ref{parametervalues}), but with $S=0.0069$ instead of $S=0.05$. The ignition values are $h(0)=12$m, $Ri(0)=\overrightarrow{Ri}=0.99$, and $R_\psi(0)=\overrightarrow{R_\psi}=0.00114$ (corresponding to $U(0)=5.58$m/s and $\psi(0)=10.64$m$^2$/s).}
 \label{Ignition}
\end{figure}

Fig.~\ref{Ignition} shows the downstream evolutions of $U$, $\psi$, and $h$ computed with the steady TEM using the parameter values as specified in P86 (see Table~\ref{parametervalues}), but with $S=0.0069$ instead of $S=0.05$.
This modified value of $S$ corresponds to $\overrightarrow{Ri}=0.99$ (from Eq.~(\ref{Ri2})). The upstream depth of the TC is set to $h(0)=12$m. Then $U(0)=5.58$m/s and $\psi(0)=10.64$m$^2$/s are obtained from the conditions $Ri(0)=\overrightarrow{Ri}=0.99$ (from Eq.~(\ref{Ribc})) and $R_\psi(0)=\overrightarrow{R_\psi}=0.00114$ (from Eq.~(\ref{Rpsibc})). It can be seen that our improved definition of the ignition values even leads to self-accelerating TCs when $\overrightarrow{Ri}$ is very close to unity, as analytically predicted.

We find that larger values of $h(0)$ also result in self-accelerating TCs, while significantly smaller values of $h(0)$ do not, in agreement with our analytical prediction that self-accelerating TCs always exist if $h(0)$ is larger than a certain minimal value. Indeed, for all values $h(0)$, $Ri$ initially increases downstream before it decreases and converges against $\overrightarrow{Ri}$ (see inset of Fig.~\ref{Ignition}). If $Ri$ exceeds unity in its initial increase, the TC rapidly dies, and this always happened if a significantly smaller value than $12$m had been chosen for $h(0)$. The upstream boundary condition $h(0)=12$m is, however, just sufficient to ensure that $Ri$ does not exceed unity in its initial increase (see inset of Fig.~\ref{Ignition}). We further confirmed our analytical prediction in Section~\ref{existence} that any value of $S$ leading to $\overrightarrow{Ri}\geq1$ never resulted in self-accelerating TCs.

We would like to emphasize that the method described above to determine the ignition values is a major improvement over the method given by P86, in which $h(0)$ is fixed, and $U(0)$ and $\psi(0)$ are obtained from $(dU/dx)(0)=(d\psi/dx)(0)=0$. If fact, for the parameter values specified above, which correspond to $\overrightarrow{Ri}=0.99$, it is impossible to obtain self-accelerating TCs with the method by P86 regardless of the value we chose for $h(0)$. Even for parameter values for which $\overrightarrow{Ri}$ is significantly smaller unity, it is remains uncertain for which values of $h(0)$ this method yields ignition. For instance, for the parameter values as specified in P86 (see Table~\ref{parametervalues}), $h(0)=1$m yields ignition and $h(0)=2$m does not.

Another major improvement of the method described above to determine the ignition values is that $h$, $U$, and $\psi$ follow the computed asymptotic behaviors (Eqs.~(\ref{psi2}-\ref{Uasym})) already at the upstream boundary (see Fig.~\ref{Ignition}), while they otherwise might require very long distances to do so and thus be extraordinarily large. The fact that $h$, $U$, $\psi$ are within a realistic range at the upstream boundary when it is determined using this improved method (e.g., for $S=0.05$, a corresponding ignition condition would be $h(0)=0.5$m, $U=0.87$m/s, and $\psi=0.01$m$^2$/s) shows that the asymptotic analysis we performed in Section~\ref{Proof} is not just a mathematical exercise, but has real-world relevance.

\section{Discussion} \label{Discussion}
This section contains two parts: a mathematical explanation for the increase of $\Delta E_{\mathrm{mod}}$ with $E_s$ in Fig.~\ref{deltaEmod} in Section~\ref{explanation} and a discussion of the question whether the TEM or FEM is more realistic in Section~\ref{comparison}.

\subsection{Relation between bed erosion and net turbulent kinetic energy production} \label{explanation}
In this section, we discuss the consequences of $\overrightarrow{\Delta E}>0$ for the turbulent kinetic energy $k$. This leads to a relation between the bed sediment erosion rate and the net production rate of turbulent kinetic energy which explains the results shown in Fig.~\ref{deltaEmod}. In order to do so, we first show that $\overrightarrow{xdk/dx}=0$ if $\overrightarrow{k}=k_c$, where $k_c$ is an arbitrary finite value. In fact, this follows from
\begin{eqnarray}
  k_c=\overrightarrow{k}=\overrightarrow{\left(\frac{xk}{x}\right)}=\overrightarrow{k}+\overrightarrow{x\frac{dk}{dx}}=k_c+\overrightarrow{x\frac{dk}{dx}}, \label{kc}
\end{eqnarray}
where we used l'Hospital's rule \citep{Chatterjee2012}. Hence, also $\Delta E=(h/U^2)dk/dx\sim x^{1/3}dk/dx\xrightarrow{x\rightarrow\infty}0$ (from Eqs.~(\ref{hasym}) and (\ref{Uasym})) if $\overrightarrow{k}=k_c$, which contradicts $\overrightarrow{\Delta E}>0$. This means that $\overrightarrow{k}=\infty$ and thus $\overrightarrow{k/U^2}=2\overrightarrow{\Delta E}/\overrightarrow{e_w}$, which follows from Eq.~(\ref{kU}) and $0<\overrightarrow{e_w}<\infty$. Hence, $\overrightarrow{e_wk/U^2}=2\overrightarrow{\Delta E}$ and $\overrightarrow{\beta k^{3/2}/U^3}=2\overrightarrow{\beta}(\overrightarrow{\Delta E}/\overrightarrow{e_w})^{3/2}$. With this knowledge, one can now calculate the limit $x\rightarrow\infty$ of Eq.~(\ref{energymixture}), analogous to what we did in Eq.~(\ref{eqstep1}) for the case $\overrightarrow{\Delta E}=0$, yielding
\begin{eqnarray}
 &&3\overrightarrow{\Delta E}+2\overrightarrow{\beta}(\overrightarrow{\Delta E}/\overrightarrow{e_w})^{3/2}=\frac{1}{2}\overrightarrow{e_w}(1-\overrightarrow{Ri})+C_D-\frac{1}{2}\overrightarrow{Ri}\,\overrightarrow{R_\psi}. \nonumber \\
 && \label{DeltaE}
\end{eqnarray}
Eq.~(\ref{DeltaE}) can be solved for $\overrightarrow{\Delta E}$ and implies that $\overrightarrow{\Delta E}<\infty$ since $0<\overrightarrow{e_w}<\infty$, $\overrightarrow{Ri}<\infty$, $C_D<\infty$, and $\overrightarrow{R_\psi}<\infty$. Hence, once one has determined the value of $\overrightarrow{\Delta E}$ from Eq.~(\ref{DeltaE}), one can compute the asymptotic profile of $k$ from $dk/dx=(U^2/h)\Delta E$ and Eqs.~(\ref{hasym}) and (\ref{Uasym}), reading
\begin{eqnarray}
 k\simeq\frac{\overrightarrow{\Delta E}}{\overrightarrow{R_\psi}}\left(\frac{\tilde g\overrightarrow{E_s}v_s}{\overrightarrow{Ri}}\right)^{2/3}x^{\frac{2}{3}} \label{kasym}
\end{eqnarray}
when inserting the values of $\overrightarrow{R_\psi}$ and $\overrightarrow{Ri}$ (Eqs.~(\ref{Rpsi}) and (\ref{Ri2})).

We wish to emphasize that the asymptotic profiles of $h$, $\psi$, $U$, and $k$ of self-accelerating TCs computed with the steady TEM (Eqs.~(\ref{Rpsi}-\ref{Uasym}) and (\ref{kasym})) are identical to the same profiles computed by the FEM if $C_D$ is replaced by $\overrightarrow{\alpha k/U^2}$ and one assumes $0<\overrightarrow{\alpha k/U^2}<\infty$. This is because the derivations in Section~\ref{bounded} remain the same in this case (see our statement in the first paragraph of Section~\ref{bounded}) and $\overrightarrow{\alpha k/U^2}=2\alpha\overrightarrow{\Delta E}/\overrightarrow{e_w}$ (see Eq.~(\ref{kU})), from which follows $0<\overrightarrow{\Delta E}<\infty$. Interestingly, the physical meaning of the quantity $\alpha k/U^2$ in the FEM is exactly that of the bed drag coefficient \citep{Parkeretal1986}. This means that physically relevant self-accelerating TCs computed with the FEM (those with finite, positive bed drag coefficient, $0<\overrightarrow{\alpha k/U^2}<\infty$) are qualitatively identical to those computed with the steady TEM since only the prefactors in the asymptotic profiles are different. We note that it can probably be shown that $0<\overrightarrow{\alpha k/U^2}<\infty$ for all self-accelerating TCs computed with the FEM.

Using the results above, we now take a look at the dimensional net production rate of turbulent kinetic energy, given by $P_{\mathrm{net}}=U^3\Delta E+e_wUk$ \citep{Parkeretal1986}. In both the steady TEM and FEM, $\overrightarrow{\Delta E}$ is independent of the erosion rate $\overrightarrow{E_s}$ since $\overrightarrow{e_w}$, $\overrightarrow{\beta}$, $C_D$ (or $\overrightarrow{\alpha k/U^2}$ in the FEM), $\overrightarrow{R_\psi}$, and $\overrightarrow{Ri}$ are independent of $\overrightarrow{E_s}$. This mean that the asymptotic dependency of $P_{\mathrm{net}}$ on $E_s$ is entirely incorporated in $U^3$ and $Uk$. From Eqs.~(\ref{Uasym}) and (\ref{kasym}), we thus learn that \textit{all} contributions to production and dissipation of turbulent kinetic energy and thus $P_{\mathrm{net}}$ are asymptotically proportional to $\overrightarrow{E_s}$. In fact, not only the dissipation due to erosion of bed sediment ($0.5RiR_\psi$) is asymptotically amplified by $E_s$, as argued by F85 and P86, but also the dissipation due to water entrainment ($0.5e_wRi$), viscous dissipation ($\beta k^{3/2}$), and turbulent kinetic energy production ($0.5e_wU^3+u_\ast^2U$). This eventually explains the increase of $\Delta E_{\mathrm{mod}}$ with $E_s$ in Fig.~\ref{deltaEmod}, which since $\Delta E_{\mathrm{mod}}=P_{\mathrm{net}}+\beta k^{3/2}$, and thus $\Delta E_{\mathrm{mod}}$ is also asymptotically proportional to $E_s$.

\subsection{An attempt to compare the TEM with the FEM} \label{comparison}
In this section, we attempt to compare the physical realism of the TEM with that of the FEM. While the previous parts of the paper dealt with the steady TEM and FEM versions proposed by F85 and P86, we now attempt to make a more general assessment since a large number of improved TEM and FEM versions have been proposed since these original studies were published. Because of this, the most meaningful way to compare the TEM with the FEM, in our opinion, is to evaluate the equation which all TEM and FEM versions, respectively, have in common and in which the TEM differs from the FEM: the closure for the bed shear stress (Eqs.~(\ref{bedshearTEM}) and (\ref{bedshearFEM}), respectively). In order to do so, we first briefly reiterate the assumptions behind these closures.

On the one hand, Eq.~(\ref{bedshearTEM}) follows from the idea that the fluid shear stress at the bed ($\rho_wu_\ast^2$) describes the streamwise component of the force applied by the fluid on the stationary bed per unit area. The main streamwise fluid force is the mean drag force, which is proportional to the square of the mean local flow velocity ($\overline{u}$). Hence, this assumption yields $u_\ast^2\propto\overline{u}(z_b)^2$, where $z_b$ denotes the vertical location of the top of the sediment bed. $\overline{u}(z_b)$ is then assumed to be roughly proportional to $U=\frac{1}{h}\int_{z_b}^{z_b+h}\overline{u}(z)\mathrm{d}z$ (the height-averaged flow velocity), which eventually yields $u_\ast^2\propto U^2$. On the other hand, Eq.~(\ref{bedshearFEM}) follows from the definition of the Reynolds stress ($u_\ast^2=-\overline{u^\prime v^\prime}(z_b)$). $-\overline{u^\prime v^\prime}(z_b)$ is first assumed to be proportional to $0.5\overline{\mathbf{u}^{\prime2}}(z_b)$ because both are turbulent correlations of dimension velocity square. Then $0.5\overline{\mathbf{u}^{\prime2}}(z_b)$ is assumed to be proportional to $k=0.5\frac{1}{h}\int_{z_b}^{z_b+h}\overline{\mathbf{u}^{\prime2}}(z)\mathrm{d}z$ (i.e., the height-averaged value of $0.5\overline{\mathbf{u}^{\prime2}}$), yielding $u_\ast^2\propto k$.

\begin{figure}
 \begin{center}
  \includegraphics[width=1.0\columnwidth]{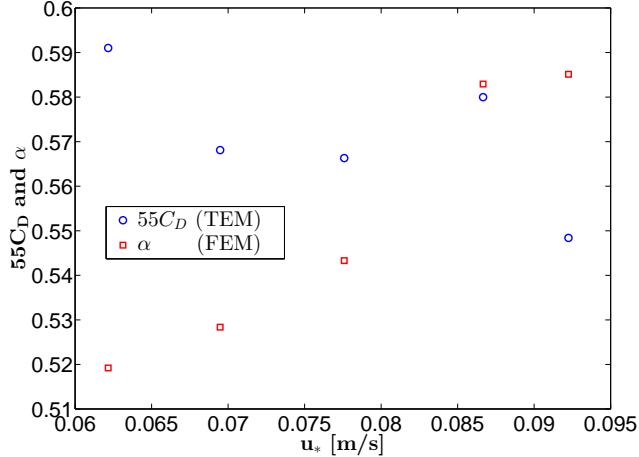}
 \end{center}
 \caption{Comparison between the TEM and FEM closures for the bed shear stress. The blue circles correspond to $55C_D=55u_\ast^2/U^2$ (see Eq.~(\ref{bedshearTEM}))and the red triangles to $\alpha=u_\ast^2/k$ (see Eq.~(\ref{bedshearFEM})), where $u_\ast$, $U$, and $k$ are obtained from the suspended load data plotted in Figs.~4 and 6-8 in \citet{Schmeeckle2014}. Constant values of $55C_D$ and $\alpha$ correspond to perfect agreement with the respective closures. $C_D$ has been multiplied by factor $55$ to make visual comparison between the closures easier.}
 \label{Closures}
\end{figure}

In the following, we compare both closures with the simulations of turbulent capacity sediment transport (using a coupled Large Eddy and Discrete Element Model) by \citet{Schmeeckle2014}. These simulations belong to the most realistic turbulent sediment transport simulations in the literature because they consider several layers of the particle bed at the scale of the particle, including particle-particle interactions as well as momentum extraction from flow due to drag on the particles. \citet{Schmeeckle2014} simulated bedload and suspended load. However, only the suspended load simulations are of interest for us since both the TEM and FEM assume that the instantaneous and thus average horizontal particle and flow velocities are the same and that there is thus no horizontal fluid drag term in the horizontal momentum balance of the fluid. In other words, the fluid shear stress, whose vertical gradient appears in this horizontal momentum balance, is assumed to be undisturbed by the presence of transported particles. However, as shown in Fig.~10 in \citet{Schmeeckle2014}, this assumption is only fulfilled in the upper parts of the flow ($z/h>0.3$, where $h$ in \citet{Schmeeckle2014} is the simulation height). Hence, the ``bed'' height ($z_b$) in the TEM and FEM is actually the height above which the local fluid shear stress is undisturbed by the presence of transported particles, which implies $z_b\approx0.3h$ for the suspended load simulations by \citet{Schmeeckle2014}. Note that $u_\ast^2(1-0.3)$ in \citet{Schmeeckle2014} thus corresponds to $u_\ast^2$ in this manuscript. Using this value of $z_b$, we used the suspended load data plotted in Figs.~4 and 6-8 in \citet{Schmeeckle2014} to compute $C_D=u_\ast^2/U^2$ and $\alpha=u_\ast^2/k$ (see Eqs.~(\ref{bedshearTEM}) and (\ref{bedshearFEM})). The values of $55C_D$ and $\alpha$ obtained in this way are plotted in our Fig.~\ref{Closures}.

It is important to note that we excluded the simulation with strongest suspended transport from this figure because, exclusively in this simulation, there is significant flow speed in the entire simulation domain (see Fig.~4 in \citet{Schmeeckle2014}). This implies a reduction of the flow resistance and thus $C_D$ which can be attributed to the finite size of the simulated system. To explain this, let us imagine, we extend the simulation domain by adding a layer of particles below $z=0$ and moving the lower simulation wall to the bottom of this added particle layer. Because the flow speed near the added particle layer is significant in the simulation with strongest suspended transport, such an extension of the simulation domain would lead to an increase of the overall horizontal drag on the particles and thus to increasing flow resistance, while such an extension of the simulation domain would have nearly no effect in the other simulations due to zero flow speed near $z=0$.

It can be seen in Fig.~\ref{Closures} that the simulations by \citet{Schmeeckle2014} seem to slightly support the use of the TEM over the FEM closure since $C_D$ varies slightly less with $u_\ast$ than $\alpha$, which seems to slightly increase with $u_\ast$. We believe that this increase is the result of turbulence damping due to density stratification, as we explain in the following. On the one hand, in order to keep the stratification stable, the flow must exert vertical drag forces on the particles which on average exactly compensate the submerged gravity forces. However, through these vertical drag forces, the flow loses turbulent kinetic energy ($k$). With increasing suspended load, the concentration and thus the submerged weight of the particles increases, resulting in a decrease of $k$. On the other hand, the vertical fluid shear stress profile and thus $u_\ast^2$ in the TEM and FEM are by definition undisturbed by the presence of transported particles (since the TEM and FEM assume that the instantaneous and thus average horizontal particle and flow velocities are the same) and thus not influenced by density stratification. Hence, $u_\ast^2/k$ increases with $u_\ast$.

We wish to emphasize, since the increase of $\alpha$ with $u_\ast$ is quite small, our comparison is just a first clue in favor of the TEM, which needs to be further supported by data in the future. Also, it is important to check in the future how both $C_D$ and $\alpha$ depend on flow parameters which remained constant in the simulations by \citet{Schmeeckle2014}, such as the particle Reynolds number ($Re_p$). 

\section{Conclusions} \label{Conclusion}
This study re-examines the steady three-equation model (TEM) for turbidity currents (TCs) by \citet{Fukushimaetal1985} (F85) and \citet{Parkeretal1986} (P86) by analytical and numerical means and compares the TEM and four-equation model (FEM) closures with predictions of recent numerical simulations. The following conclusions can be drawn from this study:
\begin{enumerate}
\item Self-accelerating TCs simulated with the steady TEM by (F85) and (P86) never violate the turbulent kinetic energy balance if a realistic value for the bed drag coefficient ($C_D$) is used (see Sections~\ref{Proof} and \ref{simulations}), which is in contrast to the nearly three decades old scientific consensus on that matter.
\item The asymptotic behaviors of self-accelerating turbidity currents have been analytically calculated in Sections~\ref{bounded} and \ref{explanation} (see Eqs.~(\ref{hasym}-\ref{Casym}) and (\ref{kasym})).
\item It is not necessary to limit the bed erosion rate ($E_s$) to allow for self-accelerating TCs, which was the motivation for the FEM, since the net production rate of turbulent kinetic energy is asymptotically proportional to $E_s$ (see Section~\ref{explanation}), even though turbulent kinetic energy is spend when suspending bed material. The physical reason behind this counter-intuitive behavior is that eroded bed sediment increases the sediment concentration and thus potential energy of the TC, which is then converted into turbulent kinetic energy downslope.
\item The steady TEM investigated in this paper has numerically stable self-accelerating solutions if and only if the Richardson number ($Ri$) is smaller than unity (supercritical flow) in the asymptotic limit ($\overrightarrow{Ri}=\lim_{x\rightarrow\infty}Ri<1$, see Section~\ref{existence}), which can be calculated by Eq.~(\ref{Ri2}). This condition is equivalent to the condition that the bed slope ($S$) must be larger than a critical value (see Eq.~(\ref{Scriteria})).
\item A novel method to determine the ignition values has been proposed (see Section~\ref{BC}). This method always leads to self-accelerating TCs if $\overrightarrow{Ri}<1$ in simulations with the steady TEM investigated in this paper, which is a major improvement over the method by \citet{Parker1982}, which often fails to do so.
\item The TEM and FEM closures for the bed shear stress are compared with state of the art simulations of suspended sediment transport using a coupled Large Eddy and Discrete Element Model (see Section~\ref{comparison}). These simulations suggests that the TEM closure (Eq.~(\ref{bedshearTEM}) performs slightly better that the FEM closure (Eq.~(\ref{bedshearFEM}).
\end{enumerate}
It is important to mention that most if not all of the conclusions summarized above can be easily generalized to more modern versions than the F85 and P86 version of the TEM investigated in this paper. For instance, unsteady self-accelerating TC solutions can be treated as fluctuations around the steady solution. Depending on the magnitude of these fluctuations, the critical asymptotic Richardson number ($\overrightarrow{Ri}$) for the existence of stable solution then will be somewhat smaller than unity. In fact, just so small that fluctuations ($Ri'$) of $Ri$ never lead to $Ri\geq1$, meaning $\overrightarrow{Ri}<1-\max Ri'$. Moreover, the conclusions summarized above indicate a strong need for studies comparing the TEM and FEM with each other in the future in order to assess which of these models is more realistic. Our study just provides a first small clue in favor of the TEM, but more investigations are needed.

\begin{acknowledgments}
The data displayed in Figs.~\ref{RpsiRi}-\ref{Closures} are available from the authors. This work was partially supported by the grants Natural Science Foundation of Zhejiang Province (LQ13E090001), Open Fund of the State Key Laboratory of Satellite Ocean Environment Dynamics (SOED1309), Natural Science Foundation of China (41376095, 41350110226, and 11402231), and Fundamental Research Funds for Central Universities of China (2013QNA4041). We thank Mark Schmeeckle most sincerely for providing us the data from his sediment transport simulations \citep{Schmeeckle2014}. We also thank editors and several reviewers for their critical and valuable comments, which led to significant improvement of our manuscript.
\end{acknowledgments}


\end{article}

\end{document}